\documentclass[%
reprint,
 superscriptaddress,
 amsmath,amssymb,
 aps,
 nofootinbib,
 prb,
 physics
]{revtex4-2}

\usepackage{booktabs}                       
\usepackage{multirow}
\usepackage{bigdelim}
\usepackage[dvipdfmx]{graphicx}
\usepackage{color}
\usepackage{dcolumn}
\usepackage{physics}
\usepackage{mathtools}
\usepackage{bm}
\usepackage{array}
\usepackage{comment}
\usepackage{cancel}
\usepackage{here}
\usepackage[dvipsnames]{xcolor}
\usepackage{mathrsfs}
\usepackage{amsmath,amsthm}
\usepackage{enumerate} 
\usepackage{empheq} 
\theoremstyle{definition}

\usepackage{subfig}

\newcommand{\mb}[1]{\mathbb{#1}}
\newcommand{\mc}[1]{\mathcal{#1}}

\newcommand{\tE}{\tilde{E}}

\newcommand{\IN}{\mb{N}}
\newcommand{\IQ}{\mb{Q}}
\newcommand{\IR}{\mb{R}}
\newcommand{\IZ}{\mb{Z}}
\newcommand{\pd}{\partial}
\newcommand{\wb}[1]{\overline{#1}}
\newcommand{\re}{{\rm e}}
\newcommand{\ri}{{\mathsf{i}}}
\newcommand{\rd}{{\rm d}}
\newcommand{\nn}{\nonumber \\}
\def\be{\begin{equation}}
\def\ee{\end{equation}}
\def\bea{\begin{eqnarray}}
\def\eea{\end{eqnarray}}

\usepackage{xcolor} 
\definecolor{RoyalBlue}{rgb}{0.255, 0.412, 0.882}
\definecolor{DeepSkyBlue}{rgb}{0.0, 0.749, 1.0}
\definecolor{Crimson}{rgb}{0.862, 0.078, 0.235}
\definecolor{ForestGreen}{rgb}{0.133, 0.545, 0.133}
\definecolor{OrangeRed}{rgb}{1.0, 0.271, 0.0}
\definecolor{Orchid}{rgb}{0.855, 0.439, 0.839}
\definecolor{Sienna}{rgb}{0.627, 0.322, 0.176}
\definecolor{Goldenrod}{rgb}{0.855, 0.647, 0.125}
\definecolor{CadetBlue}{rgb}{0.372, 0.619, 0.627}
\definecolor{CornflowerBlue}{rgb}{0.392, 0.584, 0.929}
\definecolor{RebeccaPurple}{rgb}{0.4, 0.2, 0.6}
\definecolor{Salmon}{rgb}{0.980, 0.502, 0.447}
\definecolor{HotPink}{rgb}{1.0, 0.412, 0.706}
\definecolor{Chocolate}{rgb}{0.824, 0.412, 0.118}
\definecolor{SteelBlue}{rgb}{0.275, 0.510, 0.706}
\definecolor{FireBrick}{rgb}{0.698, 0.133, 0.133}
\definecolor{bondiblue}{rgb}{0.0, 0.58, 0.71}
\definecolor{celestialblue}{rgb}{0.29, 0.59, 0.82}
\definecolor{coolblack}{rgb}{0.0, 0.18, 0.39}
\definecolor{frenchblue}{rgb}{0.0, 0.45, 0.73}
\definecolor{lapislazuli}{rgb}{0.15, 0.38, 0.61}
\definecolor{mediumpersianblue}{rgb}{0.0, 0.4, 0.65}
\definecolor{darkpowderblue}{rgb}{0.0, 0.2, 0.6}
\definecolor{darkcandyapplered}{rgb}{0.64, 0.0, 0.0}
\definecolor{darkscarlet}{rgb}{0.34, 0.01, 0.1}
\definecolor{falured}{rgb}{0.5, 0.09, 0.09}
\definecolor{darkcyan}{rgb}{0.0, 0.55, 0.55}

\def\beq{\begin{equation}}
\def\eeq{\end{equation}}
\def\bp{\begin{pmatrix}}
\def\ep{\end{pmatrix}}

\usepackage[pdftex,colorlinks,urlcolor=darkcyan,citecolor=blue,linkcolor=blue]{hyperref}

\begin{document}

\author{Jie Gu}
\affiliation{School of physics \& Shing-Tung Yau Center, Southeast University, Nanjing  211189, P. R. China}

\author{Yunfeng Jiang}
\email{jinagyf2008@seu.edu.cn}
\affiliation{School of physics \& Shing-Tung Yau Center, Southeast University, Nanjing  211189, P. R. China}
\affiliation{Peng Huanwu Center for Fundamental Theory, Hefei, Anhui 230026, China}

\author{Huajia Wang}
\email{wanghuajia@ucas.ac.cn}
\affiliation{Kavli Institute for Theoretical Sciences, University of Chinese Academy of Sciences, Beijing 100190, China}
\affiliation{Peng Huanwu Center for Fundamental Theory, Hefei, Anhui 230026, China}

\date{\today}

\title{Resurgence of $T\bar{T}$-deformed Partition Function}
\begin{abstract}
We study non-perturbative effects of torus partition function of the $T\bar{T}$-deformed 2d CFTs by resurgence. The deformed partition function can be written as an infinite series of the deformation parameter $\lambda$. We develop highly efficient methods to compute perturbative coefficients in the $\lambda$ expansion. To exemplify, the first 600 coefficients for the $T\bar{T}$-deformed free boson and free fermion are computed. Equipped with the large order perturbative data, we provide convincing numerical evidence that the $\lambda$ expansion series is asymptotic and not Borel resummable. We extract the non-perturbative contribution by resurgence and propose that they originate from new complex saddle points after analytically continuing the modular parameters in the integral representation of the partition function. The proposal is checked by comparing the predicted asymptotic behavior of the coefficients and large order perturbative data, which match nicely. The implications of these non-perturbative contributions for the Stokes phenomenon, which relates the positive and negative signs of $\lambda$, is also discussed.

\end{abstract}

\maketitle

\section{Introduction}
\label{sec:intro}
$T\bar{T}$-deformation \cite{Smirnov:2016lqw,Cavaglia:2016oda} offers us a valuable chance to study non-local quantum field theories (QFT) in depth. As an irrelevant deformation, the ultra-violet behavior of the deformed theory is significantly altered and exhibit features of gravity \cite{Cardy:2018sdv,Dubovsky:2017cnj,Dubovsky:2018bmo,Tolley:2019nmm,Hansen:2020hrs} and string theory \cite{Frolov:2019nrr,Baggio:2018gct,Caselle:2013dra,Giveon:2017nie,Callebaut:2019omt,Sfondrini:2019smd}. However, contrary to generic irrelevant deformations which are plagued by ambiguities, $T\bar{T}$-deformation is solvable. As a result, a variety of interesting physical quantities can be computed exactly in deformed theory, based on the knowledge of their undeformed counterparts.\par

An important class of solvable QFTs in 1+1 dimensions are the conformal field theories (CFTs). $T\bar{T}$-deformation of CFTs are particularly interesting due to their holographic interpretations \cite{McGough:2016lol,Guica:2019nzm,Apolo:2023ckr,Apolo:2019zai,Giveon:2017nie,Giveon:2017myj,Dei:2024sct,Benjamin:2023nts}. A fundamental quantity of a 2d CFT is the torus partition function. The $T\bar{T}$-deformed partition function has been investigated in \cite{Cardy:2018sdv,Dubovsky:2018bmo,Datta:2018thy,Aharony:2018bad,Hashimoto:2019wct,He:2020cxp}, where it was proven that the deformed partition function is modular invariant. However, unlike the deformed spectrum which takes a simple closed form,
the deformed partition function can only be given in a more formal way, as an infinite series in the deformation parameter $\lambda$, or as an integral transform of the CFT partition function \cite{Dubovsky:2018bmo,Callebaut:2019omt,Hashimoto:2019wct}. This is partly due to the complexity of the torus partition function. A direct computation of higher order contributions to the deformed partition function soon becomes very tedious. As a result, the investigation of the partition function is still preliminary. Fundamental questions such as the convergence of the $\lambda$-series is still unclear and the analytic properties of the deformed partition function is largely unexplored.

In this work, we introduce powerful tools to study the $T\bar{T}$-deformed torus partition function. Firstly, we develop highly efficient methods to compute the coefficients in $\lambda$-series, based on the theory of almost holomorphic forms (see for instance \cite{modular:2008}). This allows us to obtain perturbative results at very high orders. For example, for free CFTs (both bosonic and fermionic), we obtain analytic results up to 600 orders ! Using these large order perturbative data, we find convincing evidence that the $\lambda$-series is asymptotic. In addition, the asymptotic series is not the usual Gevrey-1 type, but of a novel type, as we will show below. To analyze the asymptotic series, we apply the resurgence method \cite{Ecalle} (see \emph{e.g.} \cite{Marino:2012zq,Sauzin:2014intro,Aniceto:2018bis} for an introduction). 
The resurgence theory allows us to explore the non-perturbative effects and analytic properties of the deformed torus partition function\footnote{See \cite{Griguolo:2022xcj,Griguolo:2022hek} for a resurgence analysis for the $T\bar{T}$-deformed 2d Yang-Mills theory.}, which turns out to be quite rich. One surprising observation is that the naive simple saddle-point does not correspond to the non-perturbative contribution obtained from large order perturbative calculations. To explain the result from large order perturbation theory, we need to take into account the influence from singularities of the analytically continued CFT partition function, which shifts the saddle-points. We obtained these new non-perturbative contributions and derive the asymptotic behaviors of the expansion coefficients, which match nicely with large order perturbation results.

\section{$T\bar{T}$-deformed torus partition function}
\label{sec:setup}
The $T\bar{T}$-deformed partition function is defined by
\begin{align}
\mathcal{Z}(\tau,\bar{\tau}|\lambda)=\sum_n e^{2\pi\ri\tau_1 RP_n-2\pi\tau_2 R \mathcal{E}_n(\lambda)}
\end{align}
where $\mathcal{E}_n(\lambda)$ is the $T\bar{T}$-deformed energy \cite{Smirnov:2016lqw,Cavaglia:2016oda}. The deformed partition function satisfies the following flow equation \cite{Cardy:2018sdv,Datta:2018thy}
\begin{align}
\label{eq:flowZ}
\partial_{\lambda}\mathcal{Z}=\left[\tau_2\partial_{\tau}\partial_{\bar{\tau}}+\frac{1}{2}\left(\partial_{\tau_2}-\frac{1}{\tau_2}\right)\lambda\partial_{\lambda}\right]\mathcal{Z}
\end{align}
with the initial condition
\begin{align}
\mathcal{Z}(\tau,\bar{\tau}|0)=Z_{\text{CFT}}(\tau,\bar{\tau})\,.
\end{align}
Here $Z_{\text{CFT}}(\tau,\bar{\tau})$ is the torus partition function of a given CFT. If we expand $\mathcal{Z}(\tau,\bar{\tau}|\lambda)$ as a formal series in $\lambda$
\begin{align}
\label{eq:pertbZ}
\mathcal{Z}(\tau,\bar{\tau}|\lambda)=\sum_{k=0}^{\infty} Z_k(\tau,\bar{\tau}) \lambda^k\,,
\end{align}
equation \eqref{eq:flowZ} leads to a recursion relation for $Z_k$
\begin{align}
\label{eq:recursionZ}
Z_{k+1} = \frac{1}{2}\left(\frac{2\tau_2}{k+1} D^{(k)}\overline{D}^{(k)} -
  \frac{k}{2\tau_2}\right) Z_k,\quad k\geq 0,
\end{align}
where $D^{(k)}$ and $\overline{D}^{(k)}$ are the Ramanujan-Serre (RS) derivatives
\begin{align}
 D^{(k)} = \frac{\partial}{\partial \tau} - \frac{\ri k}{2\tau_2},\quad
  \overline{D}^{(k)} = \frac{\partial}{\partial \bar{\tau}} + \frac{\ri k}{2\tau_2}\,.
\end{align}
Alternatively, we can write down a formal solution of the flow equation \eqref{eq:flowZ}, given by \cite{Dubovsky:2018bmo,Hashimoto:2019wct}
\begin{align}
\label{eq:integral_rep}
\mathcal{Z}(\tau,\bar{\tau},\lambda)=\frac{\tau_2}{\pi \lambda} \int_{\mathcal{H}_+} \frac{d^2 \zeta}{\zeta_2^2} e^{-\frac{|\zeta-\tau|^2}{\lambda \zeta_2}} Z_{\text{CFT}}\left(\zeta,\bar{\zeta}\right)
\end{align}
The recursion relation \eqref{eq:recursionZ} provides us an efficient way to compute higher order contributions in $\lambda$ expansion while the integral representation \eqref{eq:integral_rep} is instrumental for the saddle-point analysis.

\section{Large order perturbative results}
\label{sec:Perturb}


To compute $Z_k(\tau,\bar{\tau})$ in \eqref{eq:pertbZ}, we exploit the recursion relation \eqref{eq:recursionZ}.
The RS derivatives map a modular form of holomorphic and anti-holomorphic weights $(k,k')$ to modular forms of weights $(k+2,k')$ and $(k,k'+2)$ respectively.
As argued in \cite{Datta:2018thy}, this implies that all the $Z_k$ are modular.
Take the free boson as an example.
The undeformed partition function is
\begin{equation}
\label{eq:Z0b}
  Z_0^{\text{B}}(\tau,\bar\tau) = \frac{1}{\sqrt{\tau_2}\eta(\tau)\eta(\bar\tau)},
\end{equation}
of weights $(0,0)$.
The $n$-th order contribution $Z_n^{\mathrm{B}}$ then has modular weights $(n,n)$. The key observation is that all $Z_k^{\text{B}}$'s are elements of 
the differential ring generated by 
\begin{equation}
\label{eq:gens-b}
    \{\eta^{-1}(\tau),\; \tE_2(\tau,\bar{\tau}),\; E_4(\tau),\; E_6(\tau)\}
\end{equation}
as well as their complex conjugates.
Here $\tE_2$ is the almost holomorphic Eisenstein series defined by
\begin{equation}
    \tE_2(\tau,\bar{\tau}) = E_2(\tau) - \frac{3}{\pi\tau_2}.
\end{equation}
This ring is closed under the action of RS derivatives $D^{(n)},\wb{D}^{(n')}$, where the superscripts 
are the holomorphic and anti-holomorphic modular weights of the modular objects they act upon. 
We will omit the superscripts from now on.

In addition, it turns out that each $Z_n$ can be decomposed in terms of certain ``characters''
\begin{equation}
    Z^{\text{B}}_n = \sum_{\ell=0}^n c^{\text{B}}_{n,\ell}(\tau_2) |\chi^{\text{B}}_\ell|^2,
\end{equation}
where $\chi^{\text{B}}_\ell$ are elements of the smaller differential ring $R_{\text{B}}$ of almost holomorphic modular forms generated only by \eqref{eq:gens-b} but not their complex conjugates.
These characters are constructed as
\begin{equation}
    \chi^{\text{B}}_0 = \eta^{-1},\quad \chi^{\text{B}}_\ell = (\pi\ri)^{-\ell}D^\ell \chi^{\text{B}}_0.
\end{equation}
As $\chi^{\text{B}}_\ell$ is of modular weights $(2\ell-\frac{1}{2},0)$, the coefficient $c^{\text{B}}_{n,\ell}$ should have appropriate powers of $\tau_2$ to restore the correct modular weights of $Z^{\text{B}}_n$, and in fact one can show
\begin{equation}
    \pi^{-2\ell}\tau_2^{-2\ell+n+1/2}c^{\text{B}}_{n,\ell}(\tau_2) \in \IQ.
\end{equation}
The actual values of the coefficients $c^{\text{B}}_{n,\ell}$ can be calculated using
the recursion relation
\begin{align}
    D\wb{D}(|\chi^{\text{B}}_\ell|^2) = 
    &\pi^2|\chi^{\text{B}}_{\ell+1}|^2 - \frac{4\ell^2-2\ell-1}{8}|\chi^{\text{B}}_{\ell}|^2 \nn
    + &\left(\frac{\ell(2\ell-3)}{8}\right)^2\pi^{-2}|\chi^{\text{B}}_{\ell-1}|^2,
\end{align}
which is implied by the differential relations in the ring $R_{\text{B}}$ \cite{modular:2008} together with \eqref{eq:recursionZ}.
This allows us to develop a very efficient computer algorithm capable of calculating hundreds of terms of $Z^{\text{B}}_n$ on a normal laptop. More details can be found in the Supplementary Material (SM) \cite{SM}.

Take free fermion as a second example.
The undeformed partition function is
\begin{equation}
  \label{eq:fermion-Z0}
  Z_0^{\text{F}}(\tau,\bar{\tau}) = \left|\frac{\theta_2}{\eta}\right|
  +\left|\frac{\theta_3}{\eta}\right|
  +\left|\frac{\theta_4}{\eta}\right|,
\end{equation}
also of modular weights $(0,0)$.
The undeformed as well as the $T\bar{T}$-deformed partition function is a sum of contributions from three spin structures (A,P), (A,A), and (P,A), related to each other by simple $SL(2,\IZ)$ transformations. Therefore we only need to focus on calculations in a single spin structure, for instance, (A,P).

Similar to the free boson, the coefficients $Z_n^{(\text{A,P})}$ can also be decomposed in terms of almost holomorphic ``characters''
\begin{equation}
    Z_n^{\text{(A,P)}} = \sum_{\ell=0}^n c^{\text{(A,P)}}_{n,\ell}(\tau_2)|\chi_\ell^{\text{(A,P)}}|^2.
\end{equation}
The characters are elements of the differential ring $R_{\text{F}}^{(\text{A,P})}$ generated by
\begin{equation}
\label{eq:gens-f}
   \{ \hat{\theta}_2 = \sqrt{\theta_2/\eta},\;
    \Theta_{34} = \theta_3^4 + \theta_4^4,\;
    \tE_2(\tau,\bar{\tau}), \; E_4(\tau),\; E_6(\tau)\}.
\end{equation}
They are constructed as
\begin{equation}
\label{eq:chis}
    \chi_0^{(\text{A,P})} = \hat{\theta}_2,\quad
    \chi_\ell^{(\text{A,P})} = D^\ell \chi_0^{(\text{A,P})}.
\end{equation}
The character $\chi^{\text{(A,P)}}_\ell$ is of modular weights $(2\ell,0)$, and the coefficient $c^{\text{(A,P)}}_{n,\ell}(\tau_2)$ is such that
\begin{equation}
    \pi^{-2\ell}\tau_2^{-2\ell+n}c^{\text{(A,P)}}_{n,\ell}(\tau_2)\in\IQ.
\end{equation}
Differential relations in the ring $R_{\text{F}}^{\text{(A,P)}}$, summarised in the SM, imply the recursion relation
\begin{align}
    D\wb{D}(|\chi^{\text{(A,P)}}_\ell|^2) = 
    &\pi^2|\chi^{\text{(A,P)}}_{\ell+1}|^2 -\frac{\ell^2}{2}|\chi^{\text{(A,P)}}_\ell|^2 \nn +
    &\left(\frac{\ell(\ell-1)^2} {4}\right)^2\pi^{-2}|\chi^{\text{(A,P)}}_{\ell-1}|^2,
\end{align}
which allows efficient calculation of $c^{\text{(A,P)}}_{n,\ell}(\tau_2)$.

We calculated the coefficients $Z_n$ for the free boson up to $n=600$, and evaluate them for $\tau=\ri\tau_2$.
We find that the $\lambda$ series of the $T\bar{T}$-deformed partition function is indeed an asymptotic series, in the sense that the coefficients $Z_n$ grow factorially fast.
But these coefficients display a slightly unusual asymptotic behavior
\begin{equation}
\label{eq:Zn-asymp}
    Z_n \sim \frac{\Gamma(n+\nu)}{A^{n+\nu}}\re^{B\sqrt{\frac{n+\nu}{A}}}
    \left(c_0 + c_1 \sqrt{\frac{A}{n+\nu}} + \mc{O}(1/n)\right)\,.
\end{equation}
This asymptotic behavior can be verified through computing the auxiliary sequence $s_n = n Z_n/Z_{n+1}$ which should have the large $n$ behavior
\begin{equation}
    s_n = \frac{n Z_n}{Z_{n+1}} \sim A -\frac{B}{2}\sqrt{\frac{A}{n}} + \frac{B^2-8A\nu}{8n} +\ldots.
\end{equation}
We check that this is indeed the case, as showcased in Figs.~\ref{fig:s3b}, and find the values of the parameters to be
\begin{equation}
\label{eq:pars-b}
    A = -2\tau_2, \quad B = \ri(1+\tau_2)\sqrt{\frac{2\pi c}{3}},\quad \nu = -1,
\end{equation}
where $c=1$ is the central charge of the undeformed free bosonic CFT.
We use the generalised Richardson transform, which as explained in the SM can remove higher $\mc{O}(n^{-k/2})$ corrections, to speed up the numerical convergence.

\begin{figure}
    \centering
    \subfloat[$s_n$]{\includegraphics[width=0.7\linewidth]{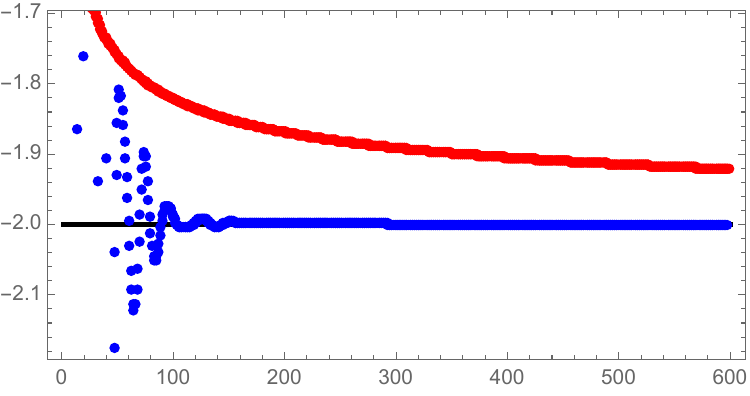}}\\
    \subfloat[$s'_n$]{\includegraphics[width=0.7\linewidth]{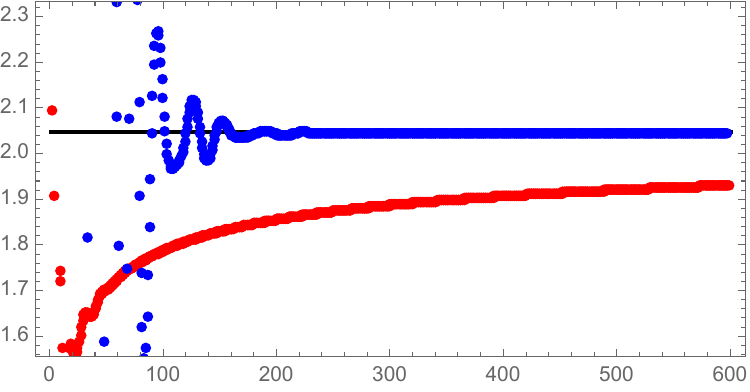}}\\
    \subfloat[$s''_n$]{\includegraphics[width=0.7\linewidth]{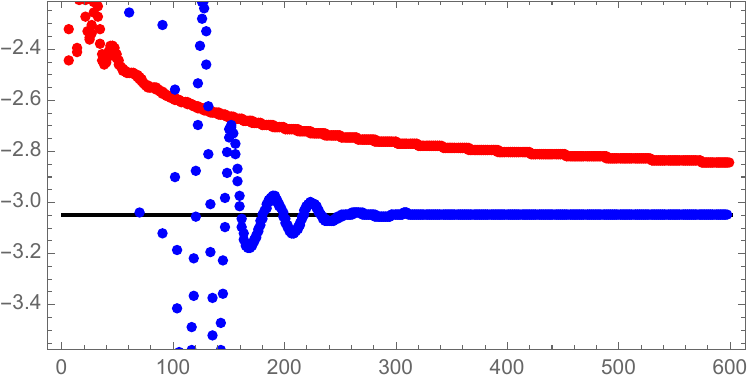}}
    \caption{The sequence of respectively $s_n$, $s'_n = \sqrt{n}(s_n-A)$, and $s''_n=n(s_n-A-(B/2)\sqrt{A/n})$ (red dots), their Richardson transforms of order-$1/2$ (blue dots), and their asymptotic values (black line), for the example of $T\bar{T}$ deformed free boson at $\tau=\ri$.}
    \label{fig:s3b}
\end{figure}




We also calculated the coefficients $Z_n$ for the free fermionic CFT up to $n=600$, and evaluated them for $\tau = \ri\tau_2$.
We find similar asymptotic behavior \eqref{eq:Zn-asymp} of the coefficients, with the same technique of auxiliary sequence $s_n$, as showcased in Figs.~\ref{fig:s3f}, and the values of the parameters are \begin{equation}
\label{eq:pars-f}
    A = -2\tau_2, \quad B = \ri(1+\tau_2)\sqrt{\frac{2\pi c}{3}},\quad \nu = -1/2,
\end{equation}
with central charge $c = 1/2$.

\begin{figure}
    \centering
    \subfloat[$s_n$]{\includegraphics[width=0.7\linewidth]{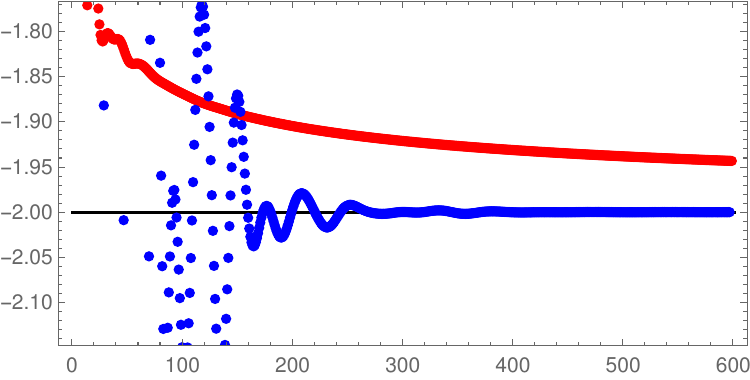}}\\
    \subfloat[$s'_n$]{\includegraphics[width=0.7\linewidth]{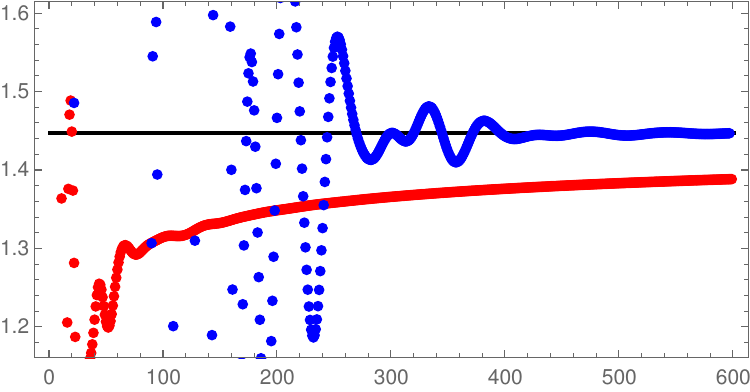}}\\
    \subfloat[$s''_n$]{\includegraphics[width=0.7\linewidth]{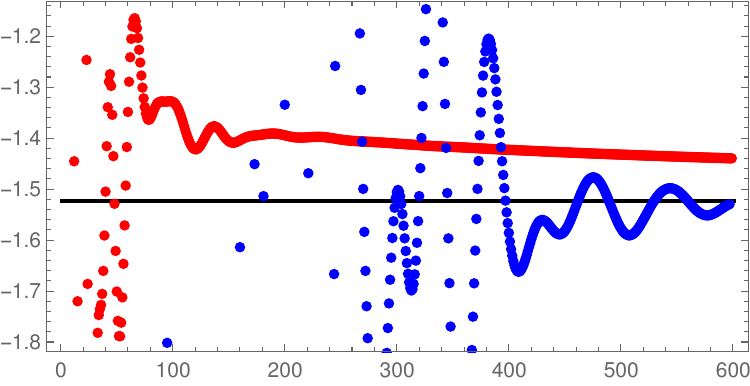}}
    \caption{The sequence of respectively $s_n$, $s'_n = \sqrt{n}(s_n-A)$, and $s''_n=n(s_n-A-(B/2)\sqrt{A/n})$ (red dots), their Richardson transforms of order-$1/2$ (blue dots), and their asymptotic values (black line), for the example of $T\bar{T}$ deformed free fermion at $\tau=\ri$.}
    \label{fig:s3f}
\end{figure}

The asymptotic behavior \eqref{eq:Zn-asymp} implies the non-perturbative corrections of the type
\begin{equation}\label{eq:np_predic}
    Z^{\text{np}} \sim \re^{-\frac{A}{\lambda}-\frac{B}{\sqrt{\lambda}}}\lambda^{-\nu}\sum_{k=0}^\infty b_k\lambda^{k/2},
\end{equation}
by the generalised large order analysis in the resurgence theory, as we explain in the SM.
We will see in the next section that $T\bar{T}$-deformed CFTs such as free boson and free fermion indeed have non-perturbative corrections of this type, with the correct parameters given in \eqref{eq:pars-b},\eqref{eq:pars-f}.

\section{Saddle-point analysis}
\label{sec:saddlepoint}
The asymptotic nature of the perturbative expansion in small $\lambda$ naturally hints at non-perturbative contributions to the deformed partition function. Most importantly, it suggests that they contribute at the order of $\mathcal{O}(e^{2\tau_2/\lambda})$. In this section we identify their origins. These should be captured from the integral representation (\ref{eq:integral_rep}). The coupling constant $\lambda$ only appears explicitly at the leading order in small $\lambda$ through the integral kernel, which admits a pair of saddle-points for $\zeta^*=\zeta^*_1+\ri \zeta^*_2$ at:
\begin{equation}
\zeta^*_1 = 0,\;\;\zeta^*_2 = \pm \tau_2
\end{equation}
By plugging them into the leading order contribution $\exp\left(-\lambda^{-1}|\zeta-\tau|^2/\zeta_2\right)$,
the ``physical" saddle at $(\zeta^*_2=\tau_2)$ contributes at the order $\mathcal{O}(\lambda^0)$, and expanding near it gives the perturbation series in (\ref{sec:Perturb}). A first guess for the origin of the non-perturbative correction is the other saddle at $(\zeta^*_2 = -\tau_2)$. However its contribution is of order $\mathcal{O}\left(e^{4\tau_2/\lambda}\right)$, inconsistent with the large order behaviour.

We point out another mechanism for new saddle-points to emerge. The free energy $F_{\text{CFT}}=-\ln{Z_{\text{CFT}}}$ of the seed CFT, though sub-leading in small $\lambda$, could be enhanced near its singularities if present. There it could play a role in the leading order saddle-point equation: 
\begin{equation}\label{eq:saddle_eqn}
\frac{1}{\lambda}\partial_\zeta \left(\frac{|\zeta-\tau|^2}{\zeta_2}\right)+\partial_\zeta F_{\text{CFT}}(\zeta,\bar{\zeta})=0
\end{equation}
and produce additional solutions. For the examples we considered, the seed partition functions do contain UV/IR divergences for $\zeta$ and $\bar{\zeta}$ independently. This is a property of rational 2d CFTs, due to its finite sum over characters. The leading order behaviour of these divergences are universal by conformal symmetry. We can approximate: 
\begin{equation}\label{eq:regge}
F_{\text{CFT}} \sim \frac{\ri\pi c }{6}\left(\zeta-\zeta^{-1}\right) -\frac{\ri\pi \bar{c}}{6}\left(\bar{\zeta}-\bar{\zeta}^{-1}\right)
\end{equation} 
if both $\zeta$ and $\bar{\zeta}$ are approaching either $0$ or $\infty$. For non-chiral CFTs with $c=\bar{c}$, and still considering only $
\tau = \ri\tau_2$, plugging (\ref{eq:regge}) into 
(\ref{eq:saddle_eqn}) gives a total of eight additional saddle-point solutions:
\begin{equation}\label{eq:regge_saddle}
\zeta^*_1 = \pm\sqrt{\frac{3\tau_2^2\pm \pi c\tau_2 \lambda}{\pi c\lambda}},\;\;\zeta^*_2 =  \pm \ri\tau_2\sqrt{\frac{3}{\pi c\lambda}}
\end{equation}
whose small $\lambda$ expansion: \footnote{There exists another branch of solutions that simply interchanges $\zeta^* \leftrightarrow \bar{\zeta}^*$, whose contributions are identical to those in (\ref{eq:regge_saddle}).} 
\begin{equation}
\zeta^* = \pm \sqrt{\frac{\pi c\lambda}{12}}+...,\;\; \bar{\zeta}^* = \pm\sqrt{\frac{12}{\pi c \lambda}} \tau_2+...
\end{equation}
are self-consistent with the approximation (\ref{eq:regge}).

These saddle-points occur in the  ``Regge" regime of the modular parameters $\zeta$ and $\bar{\zeta}$, analytically continued from the Euclidean configuration $\bar{\zeta} = \zeta^*$. \footnote{In principle $F_{\text{CFT}}$ also diverges at the images of the singularities under modular transformations, near which similar saddle-points for (\ref{eq:integral_rep}) may emerge. For free boson and free fermion, it actually can be shown that these are the only singularities. Their effects on the expansion coefficients are found to be sub-leading at large orders \cite{InProgress}. }  Furthermore, their contributions to $Z(\tau,\bar{\tau},\lambda)$, after including the $F_{\text{CFT}}$ term, take the following form:
\begin{equation}\label{eq:NP_correction}
Z(\tau,\bar{\tau},\lambda) \sim \exp\left(\frac{2\tau_2}{\lambda}+ \ri\sqrt{\frac{4\pi c}{3\lambda}}(\pm \tau_2\pm 1)\right)
\end{equation}
This matches precisely with the extracted semi-classical behavior of (\ref{eq:np_predic}), confirming the role these saddle-points play in driving the resurgence behaviour in section (\ref{sec:Perturb}). We therefore conclude that the $T\bar{T}$-deformed partition function receives non-perturbative contributions from the interplay between the $T\bar{T}$ coupling $\lambda$ and the Regge singularity of the seed partition function.

\section{Stokes' phenomena}\label{sec:stokes}
The resurgence analysis only identifies the non-perturbative corrections that can potentially contribute to $Z(\tau,\bar{\tau},\lambda)$. In general they are only present in certain wedges of the complex $\lambda$-plane. The Stokes' phenomenon refers to the discontinuous jump of the non-perturbative corrections across the wedge boundaries. Using the approximation (\ref{eq:regge}) as a toy model that captures the essential non-perturbative effects, we study the Stokes' phenomena as $\lambda$ rotates in the complex plane from the positive to negative real-axis, relating both signs of the $T\bar{T}$ coupling. We include the details in the SM. 

The physical integration contour $\mathcal{H}_+$ can be decomposed into a union of steepest descent contours through the saddle-points labelled by $m$, known as the Lefschetz thimbles $\mathcal{J}_m$ : 
\begin{equation}
\mathcal{H}_+ = \cup_{m} c_m \mathcal{J}_m
\end{equation}
Each $\mathcal{J}_m$ represents a saddle-point contribution to $Z(\tau,\bar{\tau},\lambda)$. The Stokes' phenomena is then essentially the topological change in the decomposition as $\lambda$ varies in the complex plane. 

For $\lambda>0$ and $\tau_2>\tau^c_2=\sqrt{\pi c\lambda/3}$, the contour $\mathcal{H}_+$ coincides with the Lefschetz thimble through the physical saddle-point $(\zeta^*_1=0,\zeta^*_2=\tau_2)$: 
\begin{equation}\label{eq:PF_1}
Z(\tau,\bar{\tau},\lambda) \sim e^{\frac{2\tau_2}{\lambda}-\frac{2\sqrt{\left(3\tau_2^2-\pi c\lambda\right)\left(3-\pi c\lambda\right)}}{3\lambda}}
\end{equation}
In particular, the other saddle-point contributions are absent despite being exponentially large. As $\lambda=|\lambda|e^{i\theta}$ rotates in the complex $\lambda$-plane, it can be checked that a Stokes' phenomenon is encountered at $\theta=\pi$, where the saddle-points from (\ref{eq:regge_saddle}) begin to contribute to $Z(\tau,\bar{\tau},\lambda)$. As a result of this, the partition function for $\lambda = -|\lambda|<0$ looks like:
\begin{eqnarray}\label{eq:PF_2}
Z(\tau,\bar{\tau},-|\lambda|) &\sim & e^{-\frac{2\tau_2}{|\lambda|}+\frac{2\sqrt{\left(3\tau_2^2+\pi c|\lambda|\right)\left(3+\pi c|\lambda|\right)}}{3|\lambda|}} \nonumber\\
&+& e^{-\frac{2\tau_2}{|\lambda|}+\sqrt{\frac{4\pi c}{3|\lambda|}}+...}+...
\end{eqnarray}
In contrary to (\ref{eq:PF_1}), it is non-singular and thus is well-defined for any physical values of $\tau_2>0$.

The two terms in (\ref{eq:PF_2}) point to distinct physical properties of the theory at $\lambda<0$. The first term is obtained by substituting $\lambda \to -|\lambda|$ into the partition function at $\lambda>0$. By itself this term points to complex energies similarly obtained by substituting $\lambda \to -|\lambda|$ into the spectrum at $\lambda>0$. The second term, coming from a distinct class of saddle-points, is instead compatible with imposing a UV cut-off at $\Lambda=2|\lambda|^{-1}$: 
\begin{eqnarray} 
Z(\tau,\bar{\tau},-|\lambda|) &\sim & \int^{\Lambda} dE\; e^{-\tau_2 E+ S(E)}\nonumber\\
&\sim &  e^{-\frac{2\tau_2}{|\lambda|}+S(2/|\lambda|)} +...
\end{eqnarray}
In the second line we performed an integration by parts and collected the boundary term. It matches the second term for a Cardy-like entropy: $S(E)=\sqrt{2\pi E c/3}$.   

There are two options for treating the spectral flow equation at $\lambda<0$: allowing complex energies or imposing a UV cut-off. Our analysis suggests that they could be related to distinct saddle-point contributions to the partition function via the Stokes' phenomenon. Reconciling both into a coherent result is beyond the scope of this work. One obstacle is the singularity at $\tau_2 = \tau^c_2$ in (\ref{eq:PF_1}), reminiscent of the Hagedorn singularity in string theory \cite{PhysRevLett.25.895,Fubini:1969qb,Hagedorn:1965st}. Its impact to the Stokes' phenomenon is unclear, and its resolution possibly involves all winding sectors \cite{Sathiapalan:1986db,Kogan:1987jd,OBrien:1987kzw,ATICK1988291}. They were not considered when deriving (\ref{eq:integral_rep}) in \cite{Dubovsky:2018bmo, Callebaut:2019omt, Hashimoto:2019wct} but could play important roles \cite{Benjamin2023}. We leave this to future investigations. 

\section{Discussions}
In this paper, we studied non-perturbative effects of the torus partition of $T\bar{T}$-deformed 2d CFTs. By developing highly efficient methods for computing large order perturbative results, we provide convincing evidence that the series is asymptotic. We identify novel non-perturbative contributions for the deformed partition function which matches nicely with numerical predictions.\par

Establishing the fact that the $\lambda$-series is asymptotic is a solid step towards a deeper understanding of the deformed partition function. If the series were convergent, one would expect the physics does not depend on the sign of $\lambda$ within the convergence region. However, it is known that the physical behavior is qualitatively different for the two signs of $\lambda$, no matter how small $|\lambda|$ is. In this sense, the asymptotic nature of the $\lambda$ series is expected. The appearance of asymptotic series usually signifies non-perturbative contributions. Well-known examples of such contributions in local QFTs include instantons \cite{ZinnJustin:2002ru,Marino2015:instantons} and renormalons \cite{Beneke:1998ui}, both of which give rise to Gevrey-1 type behavior for the asymptotic coefficients. Here we encounter a novel type of non-perturbative contributions whose asymptotic coefficient growth is not Gevrey-1. It is an intriguing question to find the physical interpretation of these non-perturbative contributions, which should be deeply related to the non-locality of $T\bar{T}$-deformed theories.\par

An interesting observation is that the coefficients $Z_n(\tau,\bar{\tau})$ exhibit a nice and simple structure \eqref{eq:chis}, it is tempting to ask whether the `almost holomorphic characters' $\chi_{\ell}$ have physical meanings, both in field theory and in holography \cite{Kraus:2021cwf,Datta:2021kha}. Their mathematical properties and the relations to $T\bar{T}$-deformed characters defined in \cite{Cardy:2022mhn} is yet another interesting question.\par

The efficient method for the perturbative calculation which is developed in this paper has a wide applicability to related questions. An immediate next question is performing a similar analysis for the $J\bar{T}$- and even more general $T\bar{T}+J\bar{T}+T\bar{J}$-deformations where the flow equation and integral representations of the deformed partition functions are known \cite{Aguilera-Damia:2019tpe,Hashimoto:2019wct}. Related, it would also be highly interesting to consider the torus one-point function \cite{He:2020udl} and see if there are similar non-perturbative contributions.\par

To handle asymptotic series beyond Gevrey-1, standard tool boxes in resurgence analysis should also be extended. For example, the usual Richardson transformation which one applies to extract stable values does not apply here and a generalized version is needed. Also, the asymptotic $\lambda$-series is not Borel resummable, which hinders us from obtaining closed form results from perturbative data. Is there a method which allows us to do so ? Such questions are interesting for resurgence theory and we leave them for future works.

\section*{Acknowledgment}
We thank Yang Zhang for valuable suggestion on program optimisation.
The work of J.G. is partly supported by Startup Funding no.~4007022316 and 4007022411 of Southeast University and by the NSF of China through Grant No.~6507024099.
The work of Y.J. is partly supported by Startup Funding no. 3207022217A1 of Southeast University and by the NSF of China through Grant No. 12247103.
The work of H.W. is supported by the NSF of China through Grant No. 12175238.

\bibliography{refs.bib}

\newpage

\onecolumngrid

\begin{appendix}

\begin{center}
	\textbf{{\large Supplemental Material}}
\end{center}

\section{More on perturbative methods}
\label{sc:pert}

We explain in more details the efficient algorithm that we use to calculate the higher $T\bar{T}$ coefficients $Z_n$ of free boson and free fermion.
We focus on  the example of free boson, as the case of free fermion is very similar.

\begin{table}[h!]
  \centering
  \begin{tabular}{cccccc}
    \toprule
    & $\tau_2$ & $\eta^{-1}$ & $\tE_2$ & $E_4$ & $E_6$\\
    \midrule
    weight & $(-1,-1)$ & $(-\frac{1}{2},0)$ & (2,0) & (4,0) & (6,0)\\
    \bottomrule
  \end{tabular}
  \caption{Modular weights of ring generators (as well as $\tau_2$) for free boson.}
  \label{tb:wts-b}  
\end{table}

As explained in the main text, due to the recursion relation
\begin{equation}
    \label{eq:rec}
    Z_{k+1} = \frac{1}{2}\left(\frac{2\tau_2}{k+1}D^{(k)}\bar{D}^{(k)}-\frac{k}{2\tau_2}\right)Z_k,\quad k\geq 0,
\end{equation}
higher $T\bar{T}$ coefficients $Z_n$ are calculated by repeatedly applying on $Z_0$ the Ramanujan-Serre derivatives, defined by
\begin{equation}
    D^{(k)} = \frac{\partial}{\partial \tau} - \frac{\ri k}{2\tau_2},\quad
    \bar{D}^{(k)} = \frac{\partial}{\partial \bar{\tau}} + \frac{\ri k}{2\tau_2}.
\end{equation}
The key problem is then how to calculate the actions of the RS derivatives \emph{quickly} and \emph{efficiently}.

In the case of free boson, the undeformed partition function $Z_0$ is
\begin{equation}
    \label{eq:Z0}
    Z_0 = \frac{1}{\sqrt{\tau_2}\eta(\tau)\eta(\bar{\tau})}.
\end{equation}
After repeatedly applying the RS derivation, all the $T\bar{T}$ coefficients $Z_n$ are naturally  elements of the differential ring which is generated by 
\begin{equation}
    \{\eta^{-1}(\tau),\;\tilde{E}_2(\tau,\bar{\tau}),\; E_4(\tau),\;E_6(\tau)\},
\end{equation}
and their complex conjugates, with $\tilde{E}_2$ being the famous almost holomorphic Eisenstein series defined by
\begin{equation}
    \tilde{E}_2(\tau,\bar{\tau}) = E_2(\tau) - \frac{3}{\pi\tau_2},
\end{equation}
and whose differentials are the RS derivatives $D$ and $\bar{D}$,
as is clear from the following differential relations between the generators \cite{modular:2008}
\begin{subequations}
\begin{align}
  &D\tau_2 = 0,\label{eq:Dt2}\\
  &\wb{D} \tau_2 = 0,\label{eq:Dbt2}\\
  &D\tE_2(\tau,\bar{\tau}) =
    \frac{\pi\ri}{6}\left(
    \tE_2(\tau,\bar{\tau})^2 - E_4(\tau)\right),
    \label{eq:DE2}\\
  &\wb{D}\tE_2(\tau,\bar{\tau}) =
    \frac{3\ri}{2\pi\tau_2^2},\label{eq:DbE2}\\
  &D\eta^{-1}(\tau) =
    -\frac{\pi\ri}{12}\eta^{-1}(\tau)\tE_2(\tau,\bar{\tau}),\label{eq:Deta}\\
  &DE_4(\tau) = \frac{2\pi\ri}{3}
    \left(\tE_2(\tau,\bar{\tau})E_4(\tau) - E_6(\tau)\right),\label{eq:DE4}\\
  &DE_6(\tau) = \pi\ri\left(
    \tE_2(\tau,\bar{\tau})E_6(\tau)-E_4(\tau)^2\right).\label{eq:DE6}
\end{align}
\end{subequations}
Here we suppress the superscripts of the RS derivatives for the modular weights.
Note that all the ring generators are modular, and the modular weights are summarised in Tab.~\ref{tb:wts-b}.

The first simplification we can make in the calculation is to remove $\pi,\tau_2$ by setting $\pi=\tau_2=1$ and only restore them at the end.
The reason we can remove $\tau_2$ is that
as is clear from \eqref{eq:Dt2},\eqref{eq:Dbt2},  $\tau_2$ can be treated as a constant in the differential ring so that it can be factored out and moved to the left of RS derivatives.
To restore powers of $\tau_2$ in the end, we notice that after setting $\tau_2=1$, $Z_n$ is a linear combination of 
\begin{equation}
    X_k\bar{X}_k = (D\bar{D})^{(k/2)} Z_0,\quad k=0,2\ldots,2n,
\end{equation}
with numerical coefficients, and $X_k\bar{X}_k$ have modular weights $(k,k)$.
But each component of $Z_n$ should have modular weights $(n,n)$, and the discrepancy is precisely because we have ignored powers of $\tau_2$.
As $\tau_2$ has modular weight $(-1,-1)$, we can multiply $X_k\bar{X}_k$ by $\tau_2^{k-n}$ to restore the correct modular weights.
Likewise, we can remove $\pi$ because it is also a constant in the differential ring so that it can be factored out and moved to the left of RS derivatives.
To restore powers of $\pi$ in the end, we notice that they are only introduced by the action of $D\bar{D}$.
Each time $D\bar{D}$ is applied, a factor of $\pi^2$ is introduced, so that in the end we should multiply $X_k\bar{X}_k$ by $\pi^k$.

As the next step of simplification, we observe that $Z_n$ is real, and it is always a finite sum of
absolute value squared of polynomials of the ring generators. In
other words
\begin{equation}
  Z_n = \sum_k c_k X_k \wb{X}_k =
  \sum_k c_k |X_k|^2, \quad c_k\in \IQ,
\end{equation}
where
\begin{equation}
\label{eq:XkR}
  X_k \in \IZ[\tE_2(\tau,\bar{\tau}),
E_4(\tau), E_4(\tau),\eta(\tau)^{-1}],
\end{equation}
and is of modular weight $(k,0)$, while $\wb{X}_k = X_k^*$.
Therefore instead of considering the action of $D\bar{D}$ on $X_k\bar{X}_k$, we can focus on the action of the RS derivative $D$ on the holomorphic part $X_k$.
In fact, we find the former can be expressed in terms of the latter via 
\begin{align}
  D\wb{D}(|X_k|^2)= 
  &|DX_k|^2
  +\frac{9}{4}|\pd_{\tE_2}X_k|^2 \nn+
  &\frac{9}{4}|D\pd_{\tE_2}X_k -\frac{2\ri}{3} X_k|^2 -
  \frac{9}{4}|D\pd_{\tE_2}X_k|^2 -
  \frac{k+4}{4}|X_k|^2.
  \label{eq:DDXk}
\end{align}
The calculation of derivatives of $X_k$ can also be simplified.
Because $X_k$ is an element in the polynomial ring \eqref{eq:XkR}, its RS derivative can be written as
\begin{align}
    \label{eq:DXk}
    DX_k = 
    &\frac{\ri}{6}(\tE_2^2-E_4)\pd_{\tE_2}X_k +
    \frac{2\ri}{3}(\tE_2E_4-E_6)\pd_{E4}X_k\nn
    &+\ri(\tE_2E_6-E_4^2)\pd_{E_6}X_k + \frac{\ri}{12}\eta \tE_2\pd_{\eta}X_k,  
\end{align}
and similarly
\begin{align}
  D\pd_{\tE_2}X_k = 
  &\frac{\ri}{6}(\tE_2^2-E_4)\pd^2_{\tE_2}X_k +
  \frac{2\ri}{3}(\tE_2E_4-E_6)\pd_{E4}\pd_{\tE_2}X_k\nn
  &+\ri(\tE_2E_6-E_4^2)\pd_{E_6}\pd_{\tE_2}X_k + \frac{\ri}{12}\eta \tE_2\pd_{\eta}\pd_{\tE_2}X_k,    
\end{align}
following the chain rule of the derivation.
Therefore, all the holomorphic components $X_k$s in any $Z_n$ are linear combinations of repeated actions of $D,\pd_{\tE_2}$, and $D\pd_{\tE_2}$ on $X_{-\frac{1}{2}} = \eta^{-1}$, the holomorphic part of $Z_0$.

The formula \eqref{eq:DDXk} can be further simplified.
To this end, we investigate the actions of $D,\pd_{\tE_2}, D\pd_{\tE_2}$.
Let us define the following three operators
\begin{gather}
    D_+ =D,\quad D_- = 6\,\ri\,\pd_{\tE_2},\nn
    \Delta = 2\tE_2\pd_{\tE_2} + 4E_4\pd_{E_4} + 6E_6\pd_{E_6}-\frac{1}{2}\eta^{-1}\pd_{\eta^{-1}}.
\end{gather}
Here $D+$ ($D_-$) raises (lowers) the modular weight by two and acting $\Delta$ on a modular form returns the modular weight.
Furthermore, they form a representation of the $SL(2,\IZ)$ group, as they satisfy the following commutation relations
\begin{equation}
  [D_+,D_-] = \Delta,\quad
  [\Delta,D_+] = 2D_+,\quad
  [\Delta,D_-] = -2D_-.
\end{equation}

Let us consider the following states of these operators.
Define the state $\chi_0$
\begin{equation}
    \chi_0 = X_{-\frac{1}{2}} = \eta^{-1}.
\end{equation}
It is clear that
\begin{equation}
    \Delta \chi_0 = -\frac{1}{2}\chi_0,\quad D_-\chi_0 = 0,
\end{equation}
and thus it can be regarded as a lowest weight state.
Higher weight states are constructed straightforwardly as
\begin{equation}
    \label{eq:D+l}
    \chi_\ell = D_+^\ell\chi_0.
\end{equation}
This is a modular form of weight $2\ell-\frac{1}{2}$, i.e.
\begin{equation}
    \Delta\chi_{\ell} = (2\ell-\frac{1}{2})\chi_{\ell}.
\end{equation}
More importantly, we find that 
\begin{gather}
    D_+\chi_{\ell} = \chi_{\ell+1},\quad
    D_-\chi_{\ell} = -\frac{\ell(2\ell-3)}{2}\chi_{\ell-1}.
\end{gather}
This in particular implies that
\begin{equation}
    D\pd_{\tE_2}\chi_{\ell} = \frac{1}{6\ri}D_+D_-\chi_{\ell} = -\frac{\ell(2\ell-3)}{12\ri}\chi_{\ell}.
\end{equation}
Therefore, if $X_k = \chi_{\ell}$ with $k = 2\ell-\frac{1}{2}$, the last line of \eqref{eq:DDXk} collapses to a single term, proportional to $|X_k|^2$. And the RS derivatives act on $|\chi_{\ell}|^2$ by
\begin{align}
    \label{eq:DDbarchi^2}
    D\wb{D}(|\chi_{\ell}|^2) \,=\,
    |\chi_{\ell+1}|^2 -\frac{4\ell^2-2\ell-1}{8}|\chi_{\ell}|^2 +\left(\frac{\ell(2\ell-3)}{8}\right)^2|\chi_{\ell-1}|^2.
\end{align}
As a simple example, this formula implies that
\begin{equation}    
    D\wb{D}(|\chi_{0}^2|) = |\chi_{1}|^2 + \frac{1}{8}|\chi_{0}|^2,
\end{equation}
where, following \eqref{eq:D+l}
\begin{equation}
    \chi_{1}  = -\frac{\ri}{12}\eta^{-1}\tE_2.
\end{equation}

Using the recursion relation \eqref{eq:rec} together with the initial condition \eqref{eq:Z0}, one can show that  $Z_n$ in fact consists only of $\chi_{\ell}$, $\ell=0,1,\ldots,n$, i.e.
\begin{equation}
\label{eq:characterZn}
    Z_n = \sum_{\ell= 0}^n c'_\ell |\chi_{\ell}|^2,\quad c'_\ell\in\IQ,
\end{equation}
In addition, the right hand side can be calculated very efficiently. 
The holomorphic components $\chi_\ell$ are calculated via \eqref{eq:DXk}, and the coefficients $c'_\ell$ can be calculated from \eqref{eq:DDbarchi^2}.

Note that the structure of $Z_n$ in \eqref{eq:characterZn} is reminiscent of character decomposition of the partition function in 2d CFT. So it is tempting to interpret $\chi_\ell$ as certain characters of the $T\bar{T}$-deformed partition function.

The case of $T\bar{T}$-deformed free fermion is similar.
The undeformed partition function is a sum of contributions from three spin structures,
\begin{equation}
    Z^\text{F}_0(\tau,\bar{\tau}) = Z^{(\text{A,P})}_0 + Z^{(\text{A,A})}_0+ Z^{(\text{P,A})}_0,
\end{equation}
where
\begin{subequations}
\begin{align}
    Z^{(\text{A,P})}_0 = 
    &\sqrt{\frac{\theta_2}{\eta}\frac{\bar{\theta}_2}{\bar{\eta}}},\\
    Z^{(\text{A,A})}_0 = 
    &\sqrt{\frac{\theta_3}{\eta}\frac{\bar{\theta}_3}{\bar{\eta}}},\\
    Z^{(\text{P,A})}_0 = 
    &\sqrt{\frac{\theta_4}{\eta}\frac{\bar{\theta}_4}{\bar{\eta}}}.
\end{align}    
\end{subequations}
As the first step of simplification, we notice that the spin structure (A,P) maps to (P,A), (A,A) respectively by an S-transformation, and an S-transformation followed by a T-transformation.
These relations hold also for the $T\bar{T}$-deformed partition function, so that we only have to work out the coefficients $Z_n^{(\text{A,P})}$ in  the (A,P) spin structure.

\begin{table}
  \centering
  \begin{tabular}{ccccccc}
    \toprule
    & $\tau_2$ & $\hat{\theta}_2$ & $\Theta_{34}$ & $\tE_2$ & $E_4$ & $E_6$\\
    \midrule
    weight & $(-1,-1)$ & $(0,0)$ & (2,0) & (2,0) & (4,0) & (6,0)\\
    \bottomrule
  \end{tabular}
  \caption{Modular weights of ring generators (as well as $\tau_2$ for the (A,P) spin structure of free fermion.}
  \label{tb:wts-f}  
\end{table}

The remainder of the calculation parallels exactly that of the free boson.
We can remove $\pi$ and $\tau_2$ and only restore them at the end of the calculation. 
Then the $T\bar{T}$ coefficient $Z_n^{\text{(A,P)}}$ can be  written as
\begin{equation}
    Z_n^{(\text{A,P})} = \sum_{\ell=0}^n c'_\ell|\chi_{\ell}|^2,\quad c'_\ell\in\IQ.
\end{equation}
All the holomorphic components $\chi_{\ell}$ are elements of the polynomial ring
\begin{equation}
    \IZ[\hat{\theta}_2,\Theta_{34},\tE_2,E_4,E_6].
\end{equation}
where
\begin{equation}
    \hat{\theta}_2 = \sqrt{\theta_2/\eta},\quad 
    \Theta_{34} = \theta_3^4 + \theta_4^4.
\end{equation}
The modular weights of the ring generators are summarised in Tab.~\ref{tb:wts-f}.
The differential relations are
\begin{subequations}
\begin{align}
  &D\hat{\theta}_2(\tau) = \frac{\pi\ri}{24}\hat{\theta}_2(\tau) \Theta_{34}(\tau),\\
  &D\Theta_{34}(\tau) = \frac{\pi\ri}{3}(\tE_2(\tau,\bar{\tau})\Theta_{34}(\tau)-\Theta_{34}^2(\tau) + 2E_4(\tau)),
\end{align}
\end{subequations}
together with \eqref{eq:DE2},\eqref{eq:DE4},\eqref{eq:DE6}.

More explicitly, $\{\chi_\ell\}_{\ell\geq 0}$ form a highest weight representation of $SL(2,\IZ)$ whose generators are given by the three operators
\begin{gather}
    D_+ =D,\quad D_- = 6\,\ri\,\pd_{\tE_2},\nn
    \Delta = 2\tE_2\pd_{\tE_2} + 4E_4\pd_{E_4} + 6E_6\pd_{E_6}+2\Theta_{34}\pd_{\Theta_{34}},
\end{gather}
satisfying the commutation relations
\begin{equation}
  [D_+,D_-] = \Delta,\quad
  [\Delta,D_+] = 2D_+,\quad
  [\Delta,D_-] = -2D_-.
\end{equation}
The lowest weight state is
\begin{equation}
    \chi_{0} = \hat{\theta}_2,
\end{equation}
of weight $(0,0)$.
Higher weight states are constructed as
\begin{equation}
    \chi_{\ell} = D_+^\ell \chi_{0},
\end{equation}
of weight $(2\ell,0)$.
Here $D_+$ acts by
\begin{align}
    D_+ = D = 
    &\frac{\ri}{24}\hat{\theta}_2\Theta_{34}\pd_{\hat{\theta}_2} + \frac{\ri}{3}(\tE_2\Theta_{34}-\Theta_{34}^2+2E_4)\pd_{\Theta_{34}} \nn+
  &\frac{\ri}{6}(\tE_2^2-E_4)\pd_{\tE_2} +
  \frac{2\ri}{3}(\tE_2E_4-E_6)\pd_{E4}\nn+
  &\ri(\tE_2E_6-E_4^2)\pd_{E_6},
\end{align}
again due to the chain rule, and here we have set $\pi=1$.
Finally, the coefficients $c_\ell'$ can be calculated using the following action of RS derivatives  on $|\chi_{\ell}|^2$
\begin{align}
    D\wb{D}(|\chi_{\ell}|^2) \;=\;
    |\chi_{\ell+1}|^2 -\frac{\ell^2}{2}|\chi_{\ell}|^2 + \left(\frac{\ell(\ell-1)}{4}\right)^2|\chi_{\ell-1}|^2,
\end{align}
together with the recursion relation \eqref{eq:rec}.

\section{Detailed comparison: resurgence v.s. saddle-points}
\label{SM:Stokes}
In this section, we perform a detailed comparison between the resurgence properties of the exact expansion coefficients and the saddle-point contributions. To this end, we need to first compute the predictions for the resurgence results based on the saddle-point analysis. We begin with the integral representation of the $T\bar{T}$-deformed partition function: 
\begin{eqnarray}\label{eq:integral_rep}
Z(\tau,\bar{\tau},\lambda) = \frac{\tau_2}{\pi \lambda} \int_{\mathcal{H}_+} \frac{d^2 \zeta}{\zeta_2^2} K(\zeta,\bar{\zeta},\lambda)Z_{\text{CFT}}\left(\zeta,\bar{\zeta}\right),\quad
K(\zeta,\bar{\zeta},\lambda)= \exp\left(-\frac{|\zeta-\tau|^2}{\lambda \zeta_2}\right)\,. 
\end{eqnarray}
For the purpose of explicit computations, we approximate the CFT partition function by the following simplified expression: 
\be\label{eq:toy_model}
Z_{\text{CFT}}(\zeta,\bar{\zeta}) \approx Z_{\text{TM}}(\zeta,\bar{\zeta}) = \exp\left[-\frac{\pi c \ri}{6}\left(\zeta-\zeta^{-1}\right)+\frac{\pi \bar{c}\ri}{6}\left(\bar{\zeta}-\bar{\zeta}^{-1}\right)\right]
\ee
This is a good approximation in the Lorentzian Regge regime characterized by either $\zeta \sim 1/\bar{\zeta} \to 0$ or $\bar{\zeta}\sim 1/\zeta \to 0$,  where we have identified the relevant saddle-points to be located. We comment that the divergences of this nature, \emph{i.e.} as functions of $\zeta$ and $\bar{\zeta}$ independently, are features of the individual characters. They remain features of the partition function for rational CFTs which sums over only finitely many characters. Therefore the approximation is good for rational CFTs, which we focus on in this work. We leave the analysis for chaotic CFTs to future studies. Focusing on CFTs with $c=\bar{c}$ and setting $\tau=\ri\tau_2$, the integral features an effective action of the form: 
\be\label{eq:EFFA}
Z(\tau,\bar{\tau},\lambda) \sim  \frac{\tau_2 e^{\frac{2\tau_2}{\lambda}}}{\pi \lambda} \int_{\mathcal{H}_+} \frac{d^2 \zeta}{\zeta_2^2} e^{-I_{\text{TM}}},\quad I_{\text{TM}}(\zeta_1,\zeta_2,\lambda) = \frac{1}{\lambda}\left(\frac{\zeta_1^2+\tau_2^2}{\zeta_2}+\zeta_2\right)-\frac{\pi c}{3}\left(\zeta_2+\frac{\zeta_2}{\zeta_1^2+\zeta_2^2}\right)\,.
\ee
The saddle-points of this effective action can be solved exactly. It consists of two classes of saddle-points: those produced solely by the kernel $K(\zeta,\bar{\zeta},\lambda)$ at the leading order in $\lambda$: 
\be\label{eq:saddle_1}
\zeta^*_1=0,\;\;\zeta^*_2 = \pm \sqrt{\frac{3\tau_2^2-\pi c \lambda}{3-\pi c \lambda}} \approx \pm \tau_2 +...
\ee
and those produced by an interplay between the $T\bar{T}$ kernel $K(\zeta,\bar{\zeta},\lambda)$ and the singularities of (\ref{eq:toy_model}):
\be\label{eq:saddle_2}
\zeta^*_1 =  s''\sqrt{\frac{6\tau_2^2+ 2\pi s c\tau_2 \lambda}{2\pi c\lambda}},\;\;\zeta^*_2 = \ri\tau_2 s'\sqrt{\frac{3}{\pi c\lambda}},\;\;s,s',s''=\pm 1\,.
\ee
The contributions from these saddle-points are summarized below. First of all, the contribution from the physical saddle with $\zeta^*_2\approx \tau_2$ in (\ref{eq:saddle_1}) contributes at the perturbative order $\lambda^0$:
\begin{eqnarray}
Z^{\text{pert}}(\tau,\bar{\tau},\lambda) \sim  \exp\left(\frac{2\tau_2}{\lambda}-\frac{2\sqrt{\left(3\tau_2^2-\pi c\lambda\right)\left(3-\pi c \lambda\right)}}{3\lambda}\right) \approx Z_{\text{CFT}}(\tau,\bar{\tau})+\mathcal{O}(\lambda)
\end{eqnarray}
We use the superscript ``pert'' to emphysize this is the perturbative component of the full partition function.
Regarding the saddle-points contribution from those of our interest in (\ref{eq:saddle_2}), our goal is to carry out a thorough comparison up to higher orders in large $n$ with the resurgence property of the exact perturbative expansion coefficients $Z_n$ computed from the actual CFT seed partition functions via the recursive relation.  For this reason we have computed the higher order corrections in small $\lambda$. They come from integrating over fluctuations about the saddle-points in expanding (\ref{eq:EFFA}) and including non-Gaussian interacting terms necessary for the corresponding order in $\lambda$. Their contributions are of the form: 
\bea\label{eq:saddle_contributions_1}
Z^{\text{np}}(\tau,\bar{\tau},\lambda) \sim \lambda^{\frac{1}{2}(1+\alpha)}\exp\left(\frac{2\tau_2}{\lambda}+\ri\sqrt{\frac{4\pi c}{3\lambda}}K^{s,s'}\right)\left(1+ \frac{3\ri K^{s,s'}}{8\tau_2}\sqrt{\frac{3}{2c}}\lambda^{1/2}+\mathcal{O}(\lambda)\right)
\eea
where the different saddles from the class (\ref{eq:saddle_2}) contribute with different $K^{s,s'}=s'\tau_2+ss'$ at the next-leading order. 
To be more precise we have also modified the contribution by putting in hand an additional exponent $\alpha$ that is theory dependent. It accounts for a possible additional power law factor of $\lambda^{\alpha/2}$ appended to (\ref{eq:toy_model}) in the Regge limit. For free boson we have $\alpha =1$, while for free fermion we have $\alpha =0$. These exponents can be extracted from the transformation properties of the relevant characters $\eta(\tau)$ and $\theta_{2,3,4}(\tau)$ under the inversions $\tau\to -1/\tau$: 
\be
\eta(-1/\tau) \sim \sqrt{-i\tau} \eta(\tau),\;\; \theta_{2,3,4}(-1/\tau) \sim \sqrt{-i\tau} \theta_{4,3,2}(\tau)
\ee
Notice that the expansions in the non-perturbative sectors are controlled by powers of $\lambda^{1/2}$, which slightly differ from the standard Gevrey-1 type usually encountered in QFTs.   

\begin{figure}
    \centering
   {\includegraphics[width=0.45\linewidth]
   {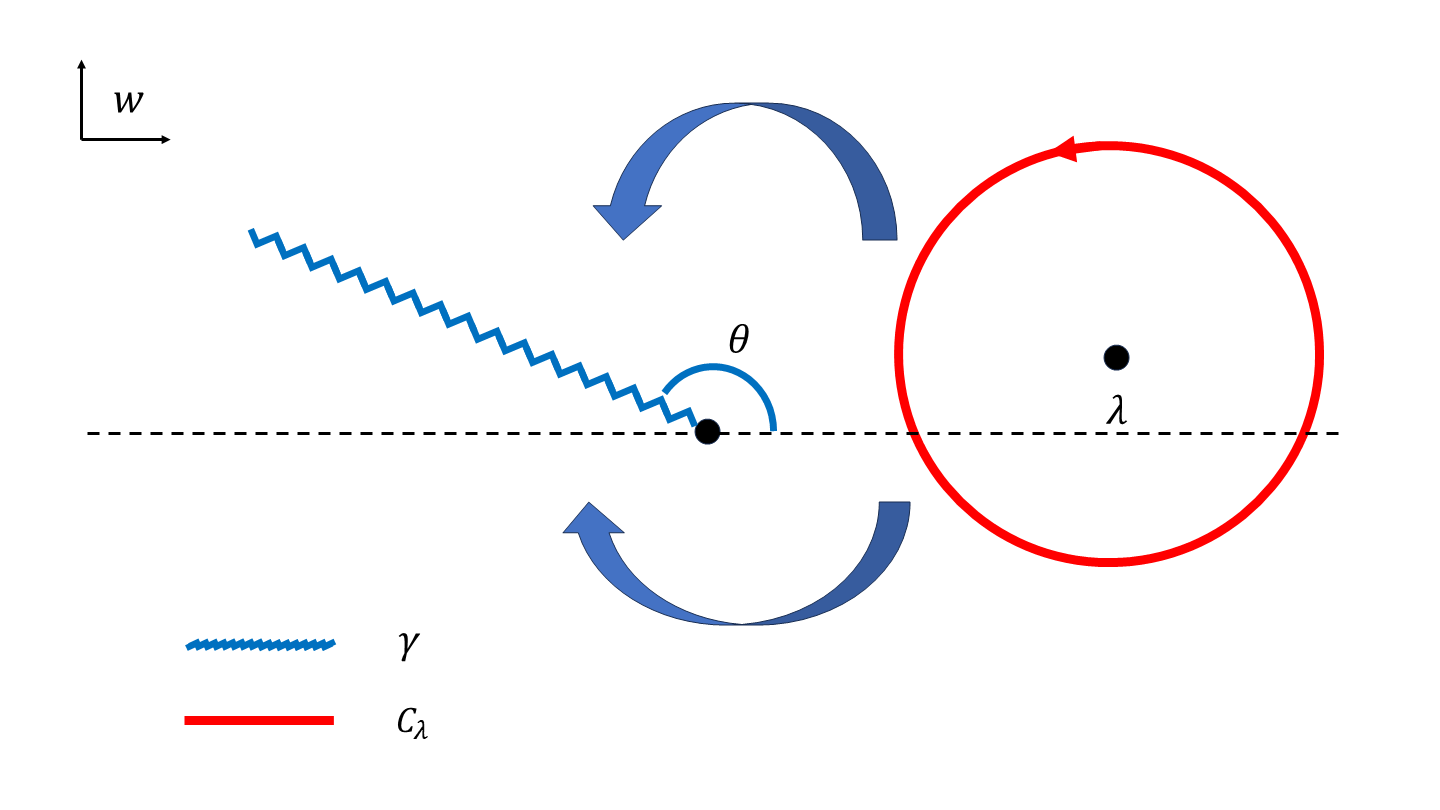}}
   {\includegraphics[width=0.45\linewidth]{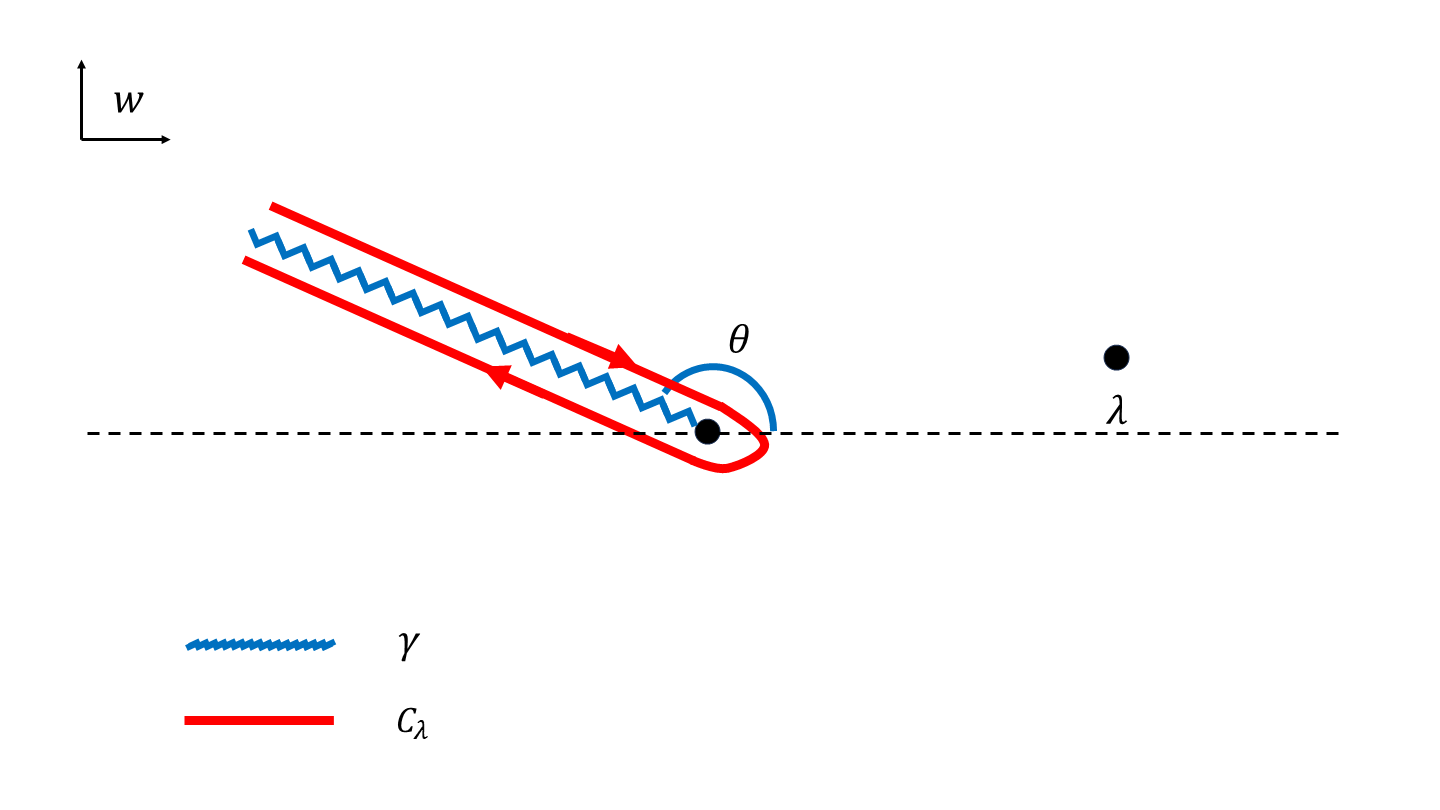}}

    \caption{Illustrated contour deformation for Eq (\ref{eq:contour_deform}). }
    \label{fig:contour_deform}
\end{figure}

Next we connect the general form of the non-perturbative contribution to the partition function that we have encountered to the resurgence properties of the perturbative expansion coefficients $Z_n$ at large orders. 
As we will recall and explain in the next section, the perturbative component $Z^{\text{pert}}$ will have discontinuity when $\lambda$ crosses Stokes' rays $\rho_\theta = \IR \re^{\ri\theta}$, and the discontinuity is given by the nonperturbative component associated to the Stokes' ray
\begin{equation}\label{eq:discZ}
    Z^{\text{pert}}(\lambda^+) - Z^{\text{pert}}(\lambda^-)  = \text{Disc}_\theta\; Z^{\text{pert}}(\lambda) = Z^{\text{np}}(\lambda) \sim \lambda^{-\nu} e^{-A/\lambda-B/\sqrt{\lambda}} \sum^{\infty}_{k=0} b_k \lambda^{k/2}
\end{equation}
with $\arg\lambda^{\pm} =\theta\pm 0^+$.
Naturally, there could be multiple Stokes curves, associated with distinct non-perturbative corrections. They are branch-cuts for the perturbative component of the partition function. To proceed, we begin by writing the perturbative partition function $Z^{\text{pert}}(\lambda)
\equiv Z^{\text{pert}}(\tau,\bar{\tau},\lambda)$ in terms of a contour integral and perform the following steps: 
\be\label{eq:contour_deform}
Z^{\text{pert}}(\lambda) 
= \frac{1}{2\pi i}\oint_{\mathcal{C}_\lambda} dw\frac{Z^{\text{pert}}(w)}{w-\lambda} = \frac{1}{2\pi i}\int_{\rho_\theta} dw \frac{\text{Disc}_\theta\;Z^{\text{pert}}(w)}{w-\lambda} 
\ee
where we first deformed the contour $\mathcal{C}_\lambda$ around $\lambda$ to wrap around the branch-cut $\rho_\theta$, making it an integral along $\rho_\theta$ of the discontinuous jump $\text{Disc}_\theta Z^{\text{pert}}(w)$, see Figure (\ref{fig:contour_deform}).

By performing small $\lambda$ asymptotic expansion on both sides of \eqref{eq:contour_deform}, this leads to an integral representation of the perturbative expansion coefficient $Z_n$: 
\be\label{eq:coeff_integral}
Z_n = \frac{1}{2\pi i}\int_{\rho_\theta}dw\; w^{-(n+1)}\; Z^{\text{np}}(w) \sim \frac{1}{2\pi i}\int^{\re^{i\theta}\infty}_0 w^{-(n+1+\nu)} e^{-A/w-B/\sqrt{w}}\sum^\infty_{k=0} b_k w^{k/2}
\ee
where $\theta$ denotes the phase of the branch-cut ray $\rho_\theta$ originating from the origin.  We can directly perform the integral for each term in the sum. Doing this yields the explicit expansion formulae: 
\bea
Z_n &\sim & \sum^\infty_{k=0}b_k \int^{\re^{i\theta}\infty}_0 dw\;e^{-A/w-B/\sqrt{w}}\;w^{-n-1-\nu+k/2}\nonumber\\
&\sim & \sum^\infty_{k=0}b_k\;\frac{\Gamma(2n+2\nu-k)}{2^{2n+2\nu-k-1}A^{n+\nu-k/2}}\;U\left(n+\nu-\frac{k}{2},\frac{1}{2},\frac{B^2}{4A}\right)
\eea
where $U$ is the Tricomi confluent hypergeometric function. A possibly more illuminating approach is to perform a saddle-point approximation of (\ref{eq:coeff_integral}) at large order, using $n$ as the large parameter. This can be done for each term in the sum over $k$ separately. To this end, it is convenient to introduce $w = x^2$,
and rewrite the integral as:
\begin{equation}
  J_n^{(k)} \sim  \int_0^{\re^{i\theta}\infty} \re^{V(x)}\rd x,\;\;
  V(x) = -\frac{A}{x^2} -\frac{B}{x} -2 n_k \log x,\quad
  n_k = n+\nu-\frac{k}{2}+\frac{1}{2}.
\end{equation}
For the resurgence analysis of $Z_n$ at large order $n$, it suffices to compute only the leading order terms, e.g. $k=0, 1$. Higher-order terms affect the higher-order in $n^{-1}$ corrections in the asymptotic behaviour of $Z_n$. The saddle point equation takes the form:
\begin{equation}
  V'(x) = \frac{2A}{x^3}+ \frac{B}{x^2}- \frac{2n_k}{x} = 0 
\end{equation}
and admits two solutions:
\begin{equation}\label{eq:large_order_saddle}
  x_{\pm} = \frac{B}{4n_k} \pm\sqrt{\frac{B^2}{16n_k^2}+\frac{A}{n_k}}.
\end{equation}
It turns out that the large order behaviour of $J^{(k)}_n$ is controlled by the saddle point $x_+$, because it is closer to the the integration contour $(0,\re^{\ri\theta}\infty)$. We can then deform the contour to pass through $x_+$ and expand near it: 
\begin{equation}
  x = x_{+} + c \delta x,\;\;\;c=\left(\frac{3A}{x_+^4}+\frac{B}{x_+^3}-\frac{n_k}{x_+^2}\right)^{-1/2}
\end{equation}
where the coefficient $c$ is chosen such that the quadratic term is a standard Gaussian: 
\be 
V(x) \approx V(x_+) - \delta x^2 + \mathcal{O}(\delta x^3)
\ee
Performing the Gaussian integrals in $\delta x$ order by order yields the following 
asymptotic formula:
\begin{equation}
  J_n^{(k)} \sim  
  \re^{\frac{B^2}{8A}}\sqrt{\frac{\pi}{3}}\;
  \frac{\Gamma\left(n+\nu-\frac{k}{2}\right)}{A^{n+\nu-\frac{k}{2}}}
  \re^{- B\sqrt{\frac{n+\nu-\frac{k}{2}}{A}}}
  \left(1 + \frac{12AB-B^3}{96A^{3/2}\sqrt{n+\nu-\frac{k}{2}}} + \mc{O}\left(1/n\right)\right)
\end{equation}
where we have used the Gaussian integral formula:
\begin{equation}
  \label{eq:gaussian}
  \int_{-\infty}^{+\infty} \rd x\;\re^{-x^2} x^{2n} = \Gamma(n+1/2).
\end{equation}
The asymptotic formula for the expansion coefficient $Z_n$ can then be assembled into:
\begin{equation}
  Z_n \sim \sum_{k=0}^\infty b_k J_n^{(k)}.
\end{equation}
For our purpose, we choose to keep corrections only up to $1/\sqrt{n}$, this involves only $k=0,1$ and we have that:
\begin{equation}
  Z_n \sim 
  \re^{\frac{B^2}{8A}}b_0\sqrt{\frac{\pi}{3}}
  \frac{\Gamma(n+\nu)}{A^{n+\nu}}
  \re^{-B\sqrt{\frac{n+\nu}{A}}}
  \left(1 +\left(\frac{b_1A^{1/2}}{b_0}+
      \frac{12AB-B^3}{96A^{3/2}}
    \right)\frac{1}{\sqrt{n+\nu-1/2}}+\mc{O}(1/n)\right).
  \label{eq:b1A}
\end{equation}
We can extract the relevant parameters by comparing \eqref{eq:saddle_contributions_1} and \eqref{eq:discZ}: 
\be\label{eq:asymp_param}
A = -2\tau_2,\;\;\;B= -i\sqrt{\frac{4\pi c}{3}} K^{s,s'},\;\;\;\nu = -\frac{1}{2}(1+\alpha),\;\;\;b_1/b_0 = \frac{3i K^{s,s'}}{8\tau_2} \sqrt{\frac{3}{2c}},\;\;\;K^{s,s'}=s'\tau_2+ss'
\ee
Notice that we have $\nu = -1$ for the free boson and $\nu = -1/2$ for the free fermion. The saddles from (\ref{eq:saddle_2}) labelled by $(s,s')$ make contributions to $Z_n$ that are equal at the leading order in large $n$. However, their contributions at the next-leading order still differ by the exponentially in $\sqrt{n}$ factor: 
\be 
Z^{s,s'}_n\propto \re^{(s'\tau_2+ss')\sqrt{2\pi cn/3\tau_2}}
\ee
As a result, what actually controls the asymptotic growth of $Z_n$ is the saddle with the most positive exponent, i.e. that with $s=s'=1$. We shall therefore focus our subsequent comparison with the prediction from this particular saddle-point. 

The actual comparisons are made with respect to the following normalized ratio between consequent coefficients: 
\be
s_n = \frac{n Z_n}{Z_{n+1}}
\ee
whose predicted large order behaviour according to (\ref{eq:b1A}) and (\ref{eq:asymp_param}) is given by:
\be\label{eq:resurgence_prediction}
s_n = A - \frac{B}{2}\sqrt{\frac{A}{n}}+ \frac{B^2-8A \nu}{8n}+... = -2\tau_2 + (1+\tau_2)\sqrt{\frac{2\pi c\tau_2}{3n}} + \frac{12\tau_2 \nu- \pi c(1+\tau_2)^2}{6 n} + ...
\ee

We compare this prediction against the $s_n$ obtained from the exact computation, for the free boson ($c=1/2,\;\nu =-1$) and free fermion ($c=1/4,\;\nu = -1/2$) respectively, and for various choices of the physical temperature $\tau_2$. To be more precise, we will extract the coefficients proportional to $n^{0}, n^{-1/2}$ and  $n^{-1}$ from the asymptotics of $s_n$ as well as two additional auxiliary sequences
\begin{equation}
    s'_n = \sqrt{n}(s_n-A),\quad s''_n = n(s_n-A-(B/2)\sqrt{A/n}),
\end{equation}
using the generalized Richardson transform method explained in the last section, and compare them with the corresponding predictions in \eqref{eq:resurgence_prediction}. We show the results for different values of $\tau$ in Fig.~\ref{fig:s3b} and Fig.~\ref{fig:s3f}. The qualitative behavior is the same, but the convergence rates to the asymptotic value depends on the explicit value of $\tau$. 

\begin{figure}
    \centering
    \subfloat[$s_n,\tau=\ri 4/5$]{\includegraphics[width=0.3\linewidth]{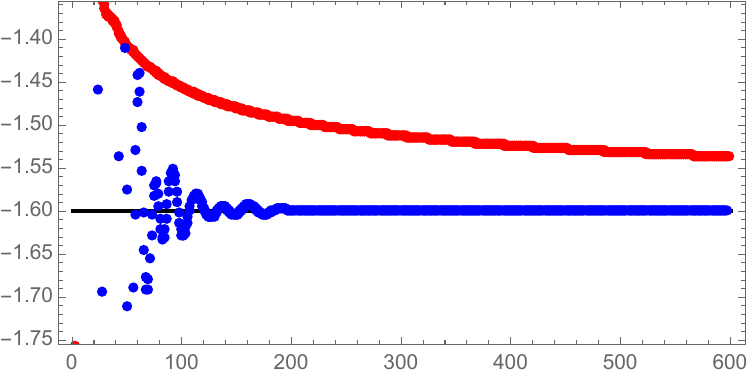}}\hspace{2ex}
    \subfloat[$s_n,\tau=\ri 9/8$]{\includegraphics[width=0.3\linewidth]{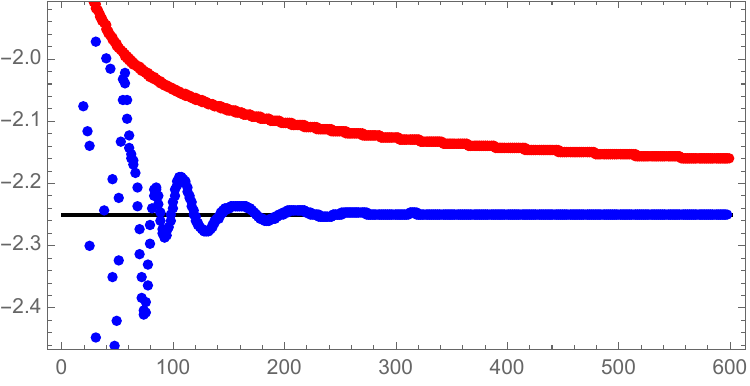}}\\
    \subfloat[$s'_n,\tau=\ri 4/5$]{\includegraphics[width=0.3\linewidth]{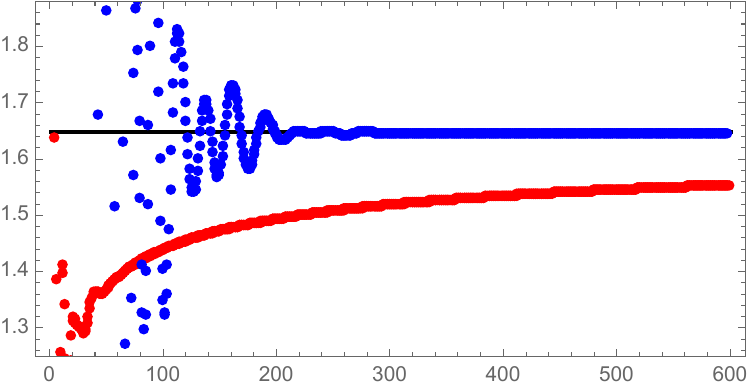}}\hspace{2ex}
    \subfloat[$s'_n,\tau=\ri 9/8$]{\includegraphics[width=0.3\linewidth]{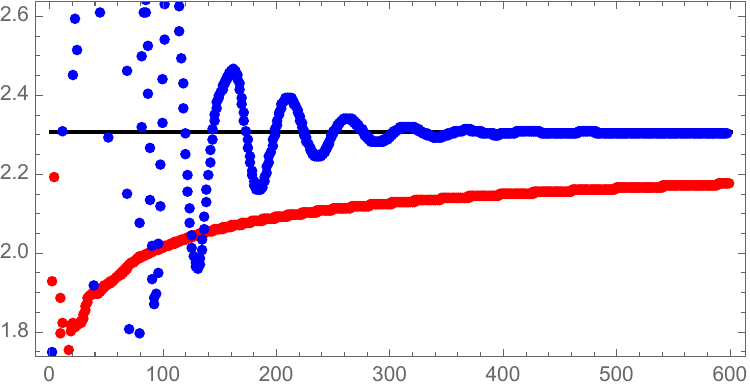}}\\
    \subfloat[$s''_n,\tau=\ri 4/5$]{\includegraphics[width=0.3\linewidth]{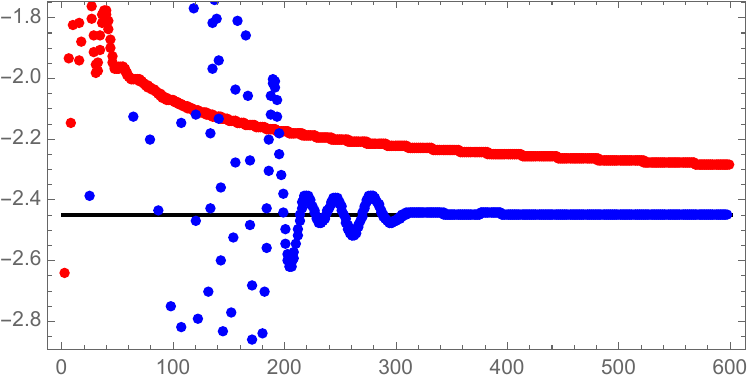}}\hspace{2ex}
    \subfloat[$s''_n,\tau=\ri 9/8$]{\includegraphics[width=0.3\linewidth]{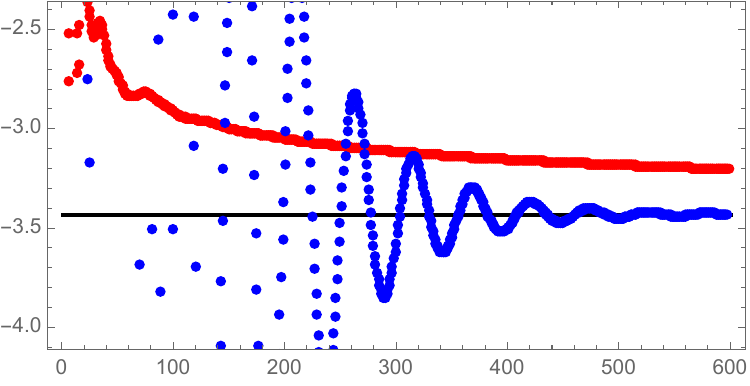}}
    \caption{The sequence of respectively $s_n$, $s'_n$, and $s''_n$ (red dots), their Richardson transforms of order-$1/2$ (blue dots), and their asymptotic values (black line), for the example of $T\bar{T}$ deformed free boson at $\tau=\ri 4/5$ and $\tau=\ri 9/8$.}
    \label{fig:s3b}
\end{figure}

\begin{figure}
    \centering
    \subfloat[$s_n,\tau=\ri 3/4$]{\includegraphics[width=0.3\linewidth]{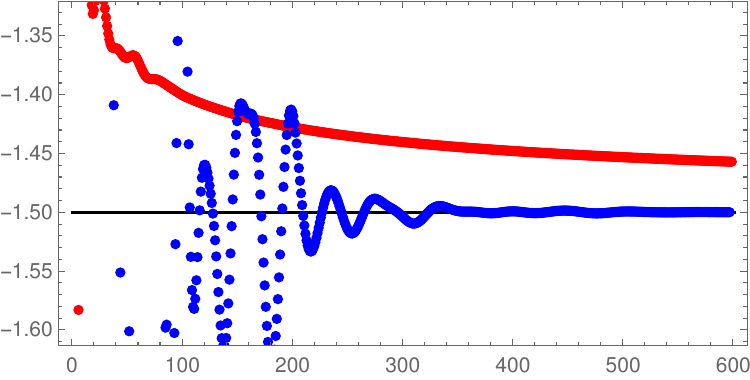}}\hspace{2ex}
    \subfloat[$s_n,\tau=\ri 4/5$]{\includegraphics[width=0.3\linewidth]
    {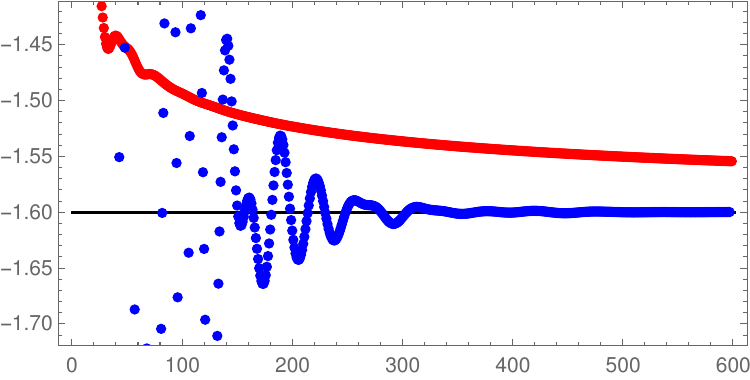}}\\
    \subfloat[$s'_n,\tau=\ri 3/4$]{\includegraphics[width=0.3\linewidth]{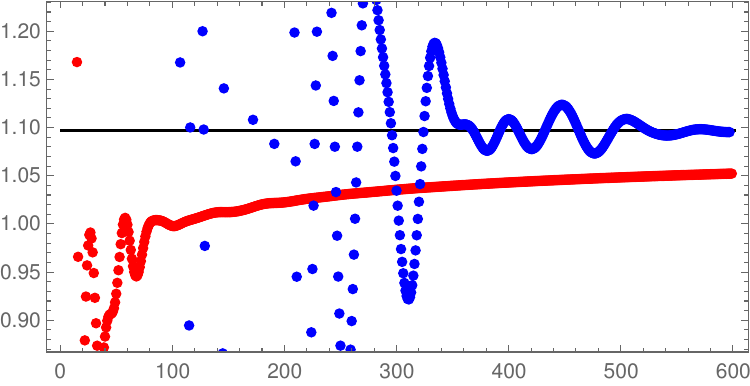}}\hspace{2ex}
    \subfloat[$s'_n,\tau=\ri 4/5$]{\includegraphics[width=0.3\linewidth]
    {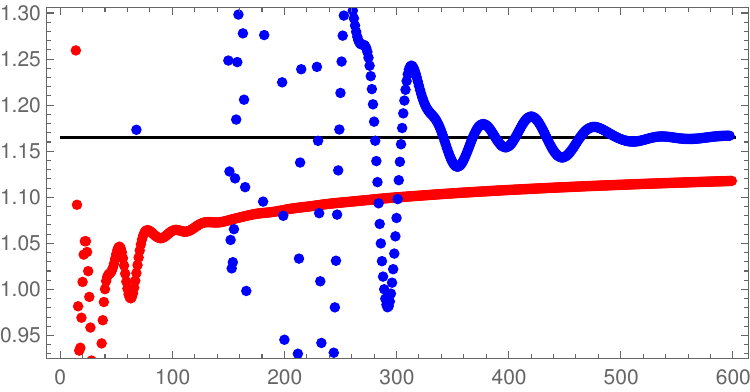}}\\
    \subfloat[$s''_n,\tau=\ri 3/4$]{\includegraphics[width=0.3\linewidth]{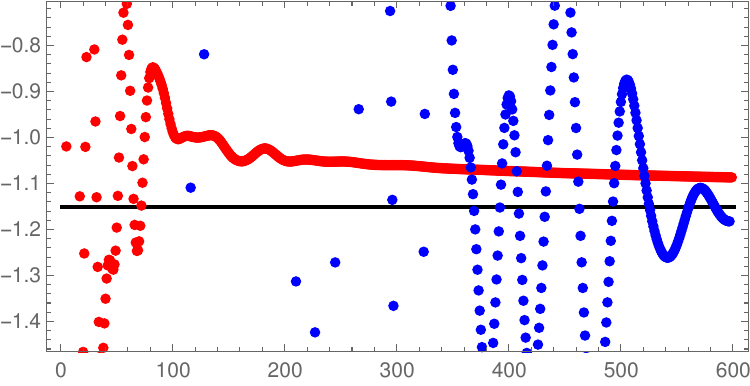}}\hspace{2ex}
    \subfloat[$s''_n,\tau=\ri 4/5$]{\includegraphics[width=0.3\linewidth]
    {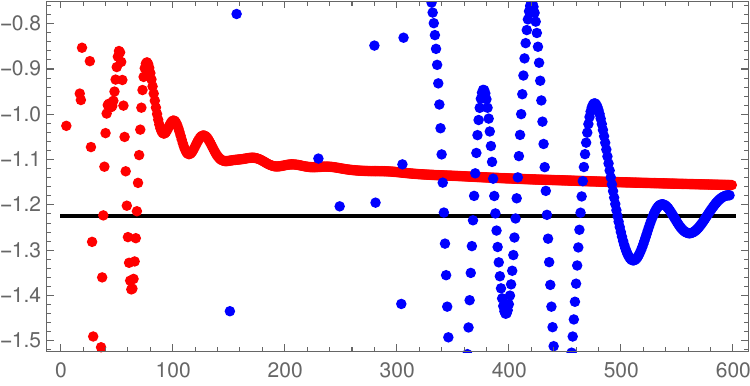}}
    \caption{The sequence of respectively $s_n$, $s'_n$, and $s''_n$ (red dots), their Richardson transforms of order-$1/2$ (blue dots), and their asymptotic values (black line), for the example of $T\bar{T}$ deformed free fermion at $\tau=\ri 3/4$ and $\tau=\ri 4/5$.}
    \label{fig:s3f}
\end{figure}


We end this section by making a few comments regarding the comparison. 
\begin{itemize}
\item The coefficients for the first three orders in the large $n$ expansion of $s_n$ match very well between the exact results and the saddle-point prediction (\ref{eq:resurgence_prediction}). This is the case for both the free boson and the free fermion CFTs. On the other hand, the exact results exhibit interesting transient oscillatory features before settling down to the predicted values. It is reasonable to suppose that they reflect the full modular-transformation properties of the seed partition functions, which are much richer than the toy model (\ref{eq:toy_model}). In particular, there are additional saddle-points of (\ref{eq:integral_rep}) that are related to (\ref{eq:saddle_2}) by modular transformations of the seed partition function. Their contributions to the expansion coefficients are found to be sub-leading at large order $n$, and oscillates with $n$ -- compatible with the qualitative features of the exact results. It is interesting to analyze these in greater details in future investigations.  

\item It is interesting to observe that it takes larger order $n$ for the exact results of the free fermions to stop oscillating and settle down to the saddle-point prediction,  compared to that of the free boson. We suspect that it is related to the ways in which the exact seed partition functions differ non-perturbatively from that of the toy model near the saddle-points. For example one can estimate such differences near the saddle-point:
\be \label{eq:regge_saddle_eg}
\zeta \approx \sqrt{\frac{\pi c}{12\lambda}}\tau_2,\;\;\bar{\zeta} \approx \sqrt{\frac{\pi c \lambda}{12}} 
\ee
For the free boson and free fermion CFTs they are controlled by: 
\bea
Z^B_0 \left(\zeta,\bar{\zeta}\right) &=& \frac{1}{\sqrt{\tau_2}\eta(\zeta)\bar{\eta}(\bar{\zeta})} \sim Z_{TM}\left(\zeta, \bar{\zeta}\right)\Big(1 + \mathcal{O}\left(q_B\right)+\mathcal{O}\left(\bar{q}_B\right)\Big),\;\;q_B=e^{2\pi i \zeta_B},\;\;\bar{q}_B = e^{-2\pi i/\bar{\zeta}_B}\nonumber\\
Z^F_0 \left(\zeta,\bar{\zeta}\right) &=& \sum^4_{i=2}\sqrt{\frac{\theta_i(\zeta)\bar{\theta}_i(\bar{\zeta})}{\eta(\zeta)\bar{\eta}(\bar{\zeta})}} \sim Z_{TM}\left(\zeta, \bar{\zeta}\right)\Big(1 + \mathcal{O}\left(q^{1/2}_F\right)+\mathcal{O}\left(\bar{q}^{1/2}_F\right)\Big),\;\;q_F=e^{2\pi i \zeta_F},\;\;\bar{q}_F = e^{-2\pi i/\bar{\zeta}_F}
\eea
where $\zeta_{B,F}$ and $\bar{\zeta}_{B,F}$ are  given by (\ref{eq:regge_saddle_eg}) by setting $c=1/2$ and $c=1/4$ respectively. We can roughly estimate their effects in the large order coefficients $Z_n$ by plugging the saddle point (\ref{eq:large_order_saddle}) for $\lambda_+ = x_+^2$ at large $n$. It is then easy to discover that: 
\be
q_B \sim \re^{-\tau_2\sqrt{\frac{\pi^3 n}{12\tau_2}}}  \ll \re^{-\tau_2\sqrt{\frac{\pi^3 n}{96\tau_2}}} = q_F^{1/2},\;\; \bar{q}_B \sim \re^{-\sqrt{\frac{\pi^3 n}{12\tau_2}}}  \ll \re^{-\sqrt{\frac{\pi^3 n}{96\tau_2}}} = \bar{q}_F^{1/2}
\ee
In other words, the effects from the deviation between the exact results and the toy model decay much faster in the free boson than that in the free fermion. This may provide an explanation for the slower convergence of the exact results to the predicted values in the free fermion. 

\item In principle, the comparison can be  extended further to the higher order coefficients of $s_n$ in the large $n$ expansion. For example, one can compare the $n^{-3/2}$ term in $s_n$, whose predicted coefficient involves $b_0/b_1$ in (\ref{eq:asymp_param}). However, at this order we found that the exact results do not match the saddle-point prediction. To explain this mismatch, one can show that near (\ref{eq:saddle_2}) the approximation (\ref{eq:toy_model}) is accurate up to order $\mathcal{O}(\lambda^{1/2})$ compared to the exact CFT partition function. Corrections of this order need to be considered in order to push the match further. One can further show that such corrections begin to affect $s_n$ precisely at the order $n^{-3/2}$. The mismatch therefore has a well-understood origin and fix: we simply need to improve the approximation (\ref{eq:toy_model}) to higher orders -- a task we shall leave for future investigations. 
\end{itemize}

In conclusion, the comparison we have performed provides strong evidence that the resurgence properties we observe in the exact expansion coefficients are indeed controlled by the saddle-points of the nature (\ref{eq:saddle_2}). 

  
\section{Stokes' phenomenon}
\label{app:stokes}

In this section, we study the Stokes' phenomenon associated with the saddle-points we have identified from the resurgence analysis. Our goal for studying the Stokes' phenomenon is as follows. In general, the non-perturbtive contribution from a particular saddle-point does not necessarily contribute for all complex values of the coupling constant $\lambda$. In which region of the complex $\lambda$-plane is it present is determined by studying the Stokes' phenomenon. This is important for analyzing the $T\bar{T}$-deformed theory with both $\lambda>0$ and $\lambda<0$ -- they are related by a rotation of $\lambda$ in the complex plane. Based on its effectiveness in capturing the non-perturbative effects, we shall use the approximation (\ref{eq:toy_model}) for analyzing the Stokes' phenomena.  

We begin with a brief recount of the basic ingredients that underlie the Stokes' phenomenon. A prototypical setting involves the integral of the form: 
\be\label{eq:Stokes_example}
F(k) = \int_{\gamma} dx\; e^{-k I(x)}
\ee
where $k$ is in general a complex parameter with large modulus. In order for the integral to be well-defined, the integration contour $\gamma$ has to asymptote towards the converging region where $\text{Re}\; k I(x) \to \infty$. The integral can be approximated by summing over contributions from the saddle-points $x_m$ satisfying: $I'(x_m) = 0$. In particular, for any given saddle one can construct the steepest descent contour $\mathcal{J}_m$ passing through $x_m$, known as the Lefschetz thimble. For one-dimensional integral $\mathcal{J}_m$ is characterized by the conditions: 
\be\label{eq:Lefschetz_def}
\text{Im}\; k I(x) = \text{Im}\; k I(x_m),\;\; \text{Re}\; k I(x) \geq \text{Re}\; k I(x_m),\;\;\;x\in \mathcal{J}_m
\ee
For higher dimensional integrals, the Lefschetz thimbles are defined as being generated by the gradient flows of the real part $h(x)=\text{Re}\;k I(x)$ \cite{fedoryuk:1977saddle}. In general, for a $n$ real-dimensional integral defined in $n$ complex-dimensions, the Lefschetz thimbles $\mathcal{J}_m$ are $n$ real-dimensional submanifolds of $\mathbb{C}^n$ through $x_m$. The conditions (\ref{eq:Lefschetz_def}) are still satisfied on $\mathcal{J}_m$, but are not sufficient to define $\mathcal{J}_m$ \cite{ursell:1980integrals,kaminski:1994exponentially}. The contribution from a particular saddle-point $x_m$ can therefore be written canonically as: 
\be
F_m(k) = \int_{\mathcal{J}_m} dx\; e^{-k I(x)}
\ee
In the large $k$ limit, it admits an asymptotic expansion
\begin{equation}
    F_m(k) \sim \frac{e^{-k I(x_m)}}{\sqrt{kI''(x_m)/2}} \left(1+\cdots\right)
\end{equation}
where ... denotes higher order corrections about this saddle-point.
For a given saddle-point $x_m$, whether it contributes to the integral (\ref{eq:Stokes_example}) depends on the integration contour $\gamma$. In generally it can be decomposed into a union of the Lefschetz thimbles \cite{pham:1967introduction}: 
\be
\gamma = \cup_m c_m \mathcal{J}_m
\ee
where the coeffcients $c_m$ take values in $(0,\pm 1)$ depending on whether it appears in the decomposition and its relative orientation with respect to $\gamma$. The integral can now be written formally as a sum of the saddle-point contributions: 
\be\label{eq:saddle_point_sum}
F(k) = \sum_m \; c_m F_m(k) \sim \sum_{m} c_m\;  \frac{e^{-k I(x_m)}}{\sqrt{kI''(x_m)/2}} \left(1+\cdots\right)
\ee
where $\sim$ denotes asymptotic expansion in $1/k$ near each saddle point. The full sum over these series through (\ref{eq:saddle_point_sum}) is called the trans-series. 

We emphasize that the definition of the Lefschetz thimbles depends on the coupling constant $k$, in particular on its phase $k = e^{i\theta}|k|$. Because of this, as $k$ rotates in the complex plane, the Lefschetz thimbles deform accordingly as submanifolds. The decomposition coefficients $c_m$ also changes, but only by integers when the Lefschetz thimble undergoes a topological change. There will be critical values for the phase $\theta$ across which the coefficients jump discontinuously, \emph{e.g.} $c_m = c_m \pm 1$. As a result, the saddle-point contributions to the integral (at fixed integration contour $\gamma$) also jump according to (\ref{eq:saddle_point_sum}). This is called the Stokes' phenomena.  The critical phase $\theta_m$ across which $c_m$ jumps describes a ray in the complex $k$-plane, that marks a edge of the wedge inside which the saddle-point contribution $F_m$ appears in (\ref{eq:saddle_point_sum}). It is the Stokes' ray associated with the saddle-point $x_m$. 

Without loss of generality, we can take the integration contour $\gamma$ to coincide with a particular Lefschetz thimble $\mathcal{J}_0$ through the saddle point $x_0$ at say $k=|k|>0$, and consider the Stokes' pheonomenon associated with another saddle $x_m$. For one-dimensional integral, it can be described geometrically in terms of when the Lefschetz thimble $\mathcal{J}_0$ passes through $z_m$, so that both $z_m$ and $z_0$ lie on $\mathcal{J}_0$. When this happens $\mathcal{J}_0$ usually takes a sharp turn at $z_m$.
It is easy to see that this happens when: 
\be\label{eq:Stokes_pheno}
\text{Im}\;k I(x_0) = \text{Im}\; k I(x_m),\;\;\; \text{Re}\;k I(x_0)< \text{Re}\;k I(x_m)
\ee
Solutions to (\ref{eq:Stokes_pheno}) specifies a set of phases $\theta_m$ for $k$, \emph{i.e.} rays of $k$ -- they are the Stokes' rays. When $k$ crosses a Stokes' ray, say, associated to $z_m$, 
$\mathcal{J}_0$ undergoes a change in topology
\begin{equation}
    \mathcal{J}_0 \rightarrow \mathcal{J}_0 + \mathcal{J}_m,
\end{equation}
which implies that the value of $F_m(k)$ has a discontinuity
\begin{equation}
    \label{eq:DiscF}
    \text{Disc}_{\theta_m}F_0(k) := F_0(k^+) - F_0(k^-) = F_m(k), 
\end{equation}
with $k^{\pm} = |k|\re^{\ri(\theta_m\pm0^+)}$.

For higher dimensional integrals, the geometric description of the Stokes' phenomenon becomes obscure. However, it can be shown that the Stokes' phenomenon remains a real co-dimension one event, i.e. it can be generically encountered by tuning only one real-parameter, e.g. the phase $\theta$ of $k$ \cite{howls:1997hyper}. When this happens, (\ref{eq:Stokes_pheno}) must still be satisfied. Technically, for higher dimensional integrals (\ref{eq:Stokes_pheno}) is only a necessary condition -- there are additional uncertainties due to the possible non-trivial topology of the Lefschetz thimbles, \emph{e.g.} they may contain multiple Riemann-sheets. In our analysis, this is settled-down by the fact that we have actually observed the resurgence effects of the saddle-points (\ref{eq:saddle_2}) in the the exact perturbative coefficients. It dictates that the corresponding Stokes' phenomenon does happen, we simply need to determine when. For this purpose, the condition (\ref{eq:Stokes_pheno}) is sufficient for the two-dimensional integral (\ref{eq:integral_rep}). 

\begin{figure}
    \centering
   {\includegraphics[width=0.33\linewidth]
   {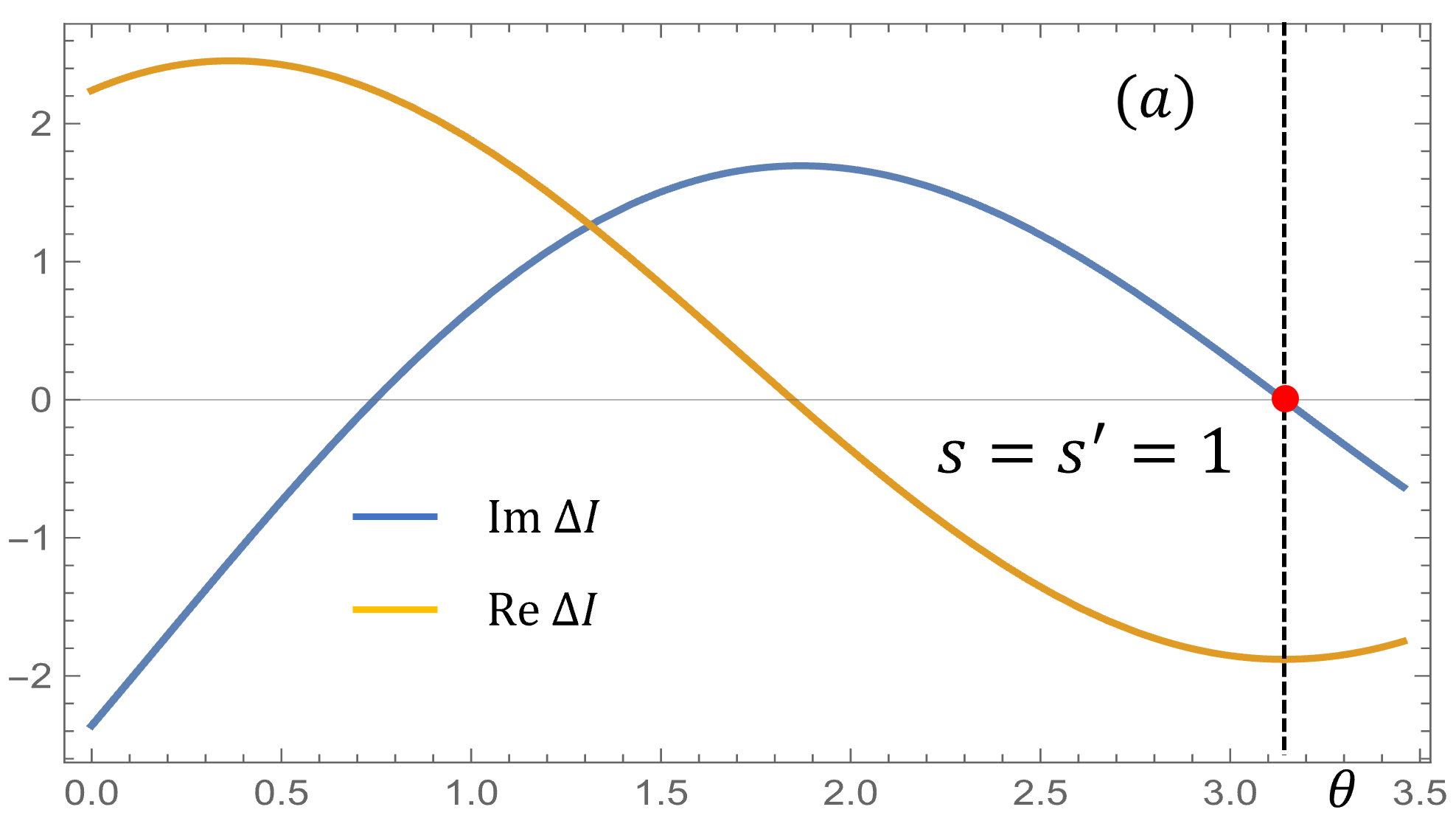}}
   {\includegraphics[width=0.33\linewidth]{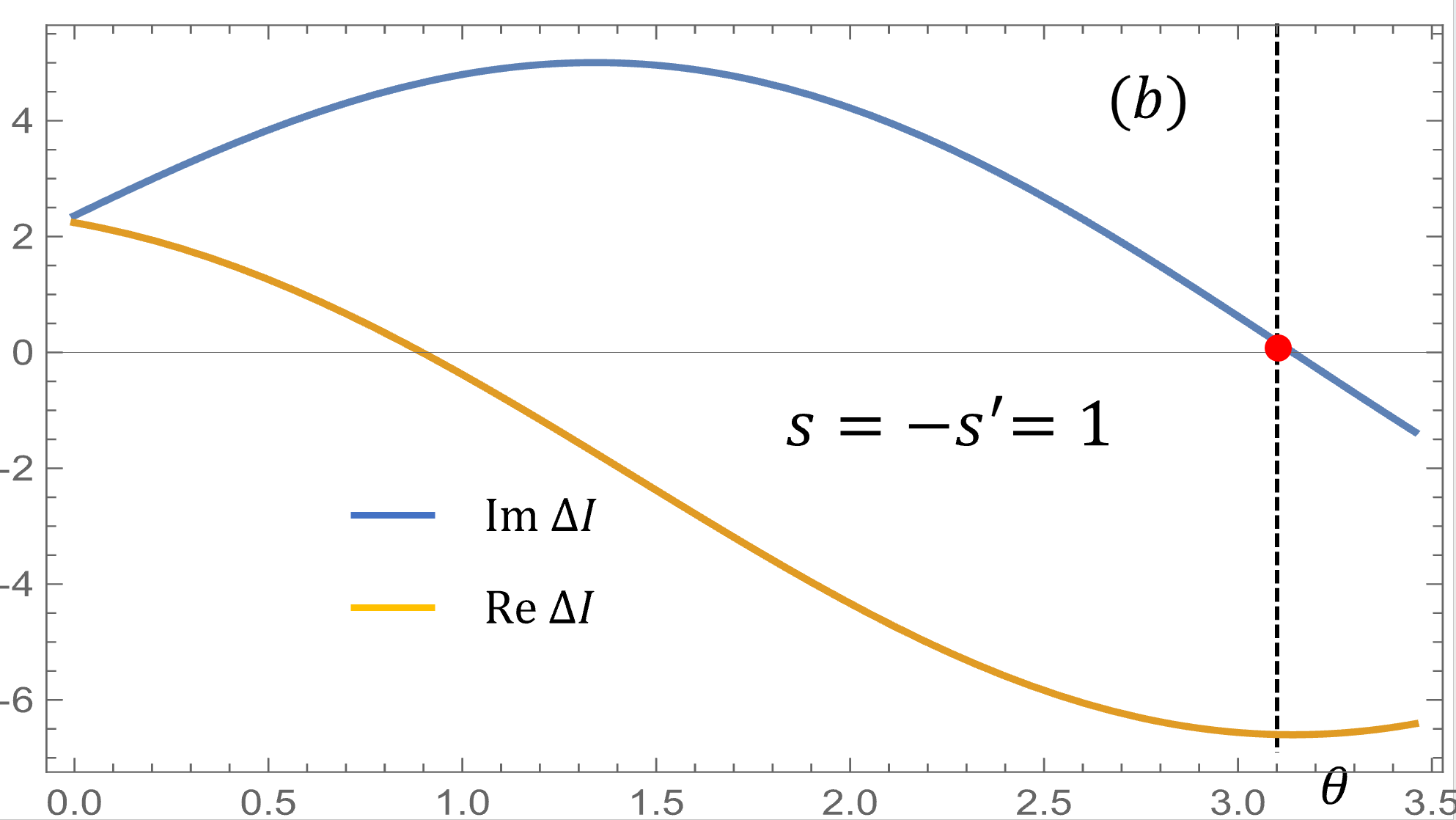}}
   {\includegraphics[width=0.33\linewidth]{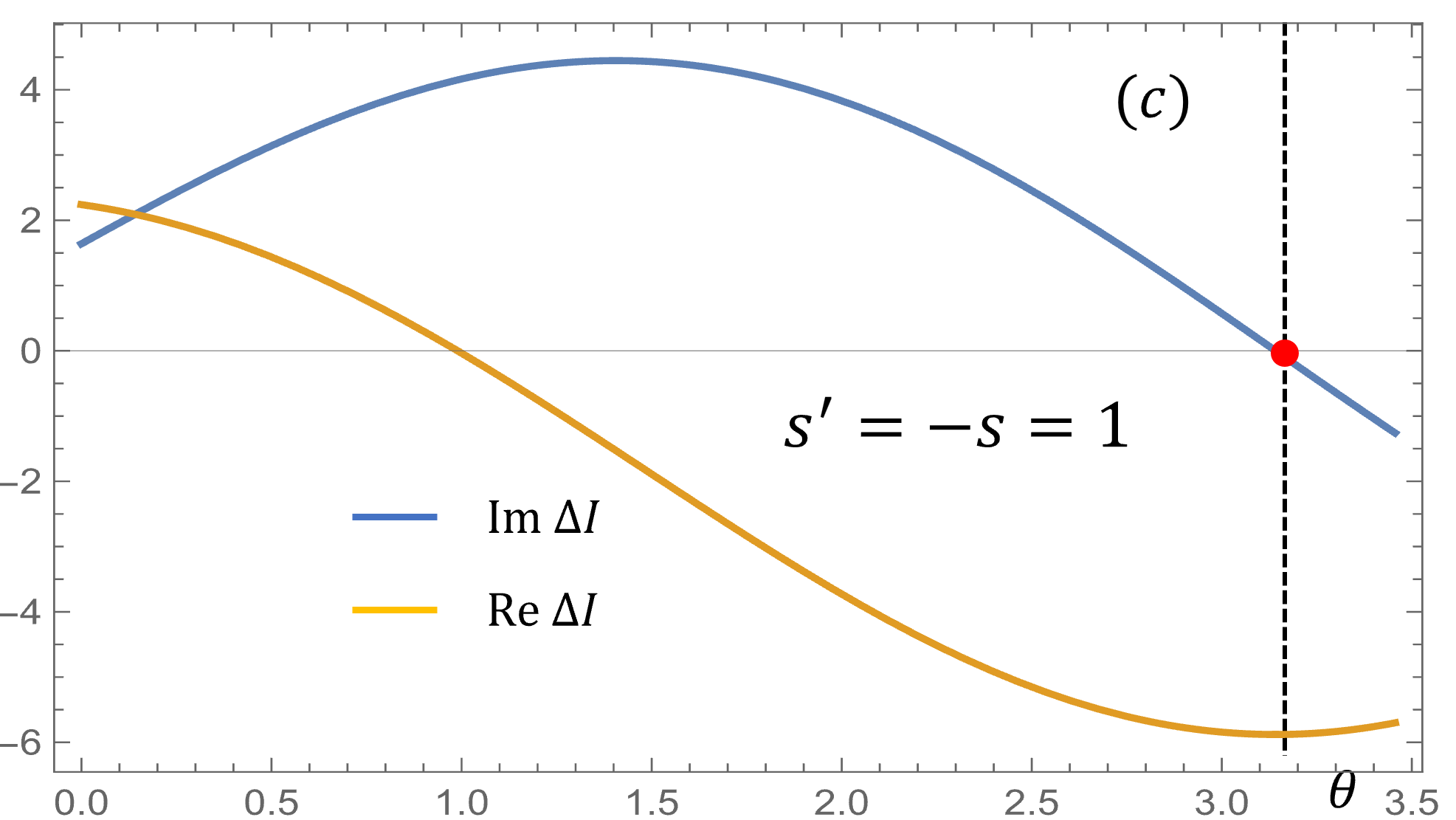}}
   {\includegraphics[width=0.35\linewidth]{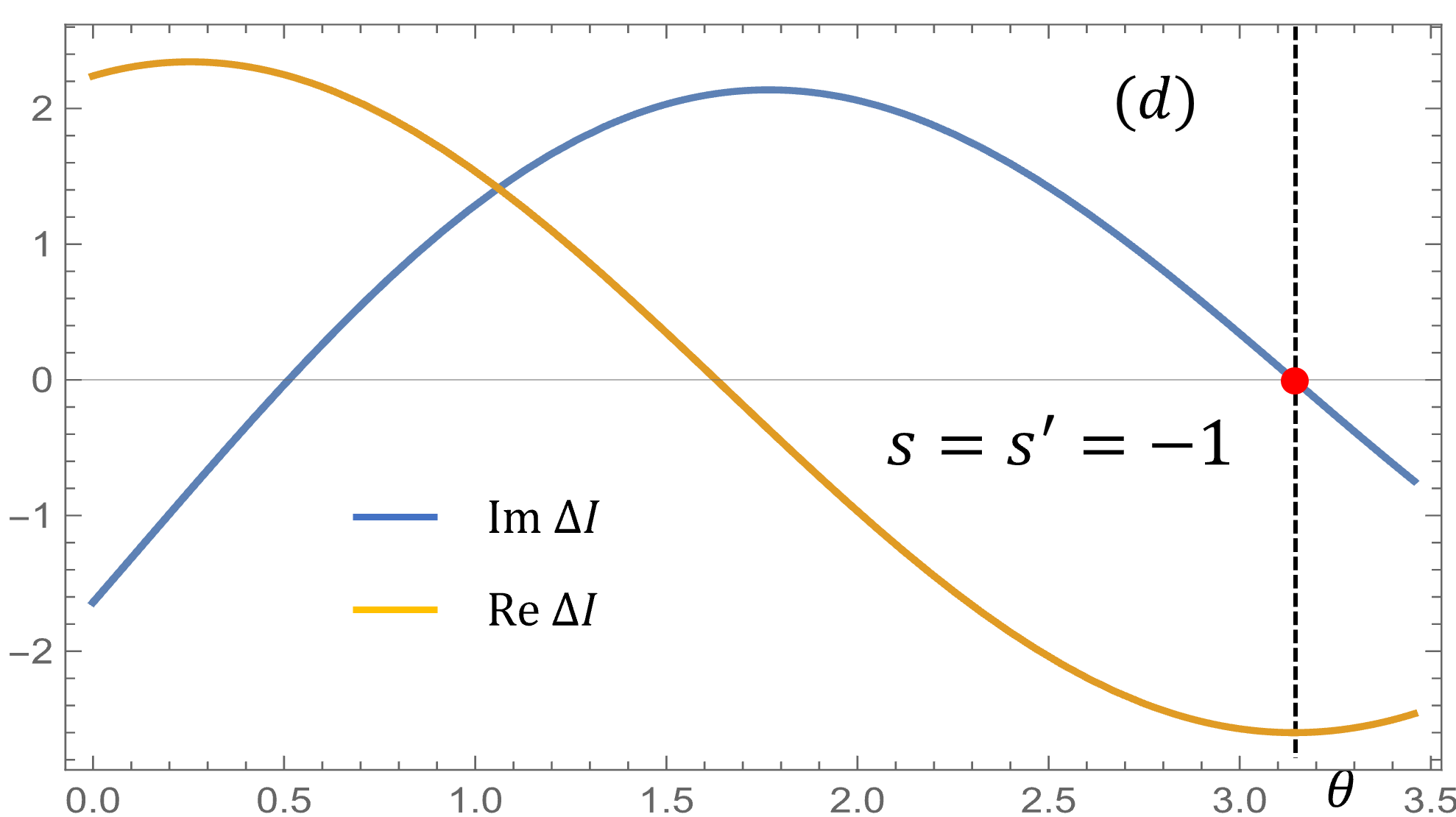}}
    {\includegraphics[width=0.35\linewidth]{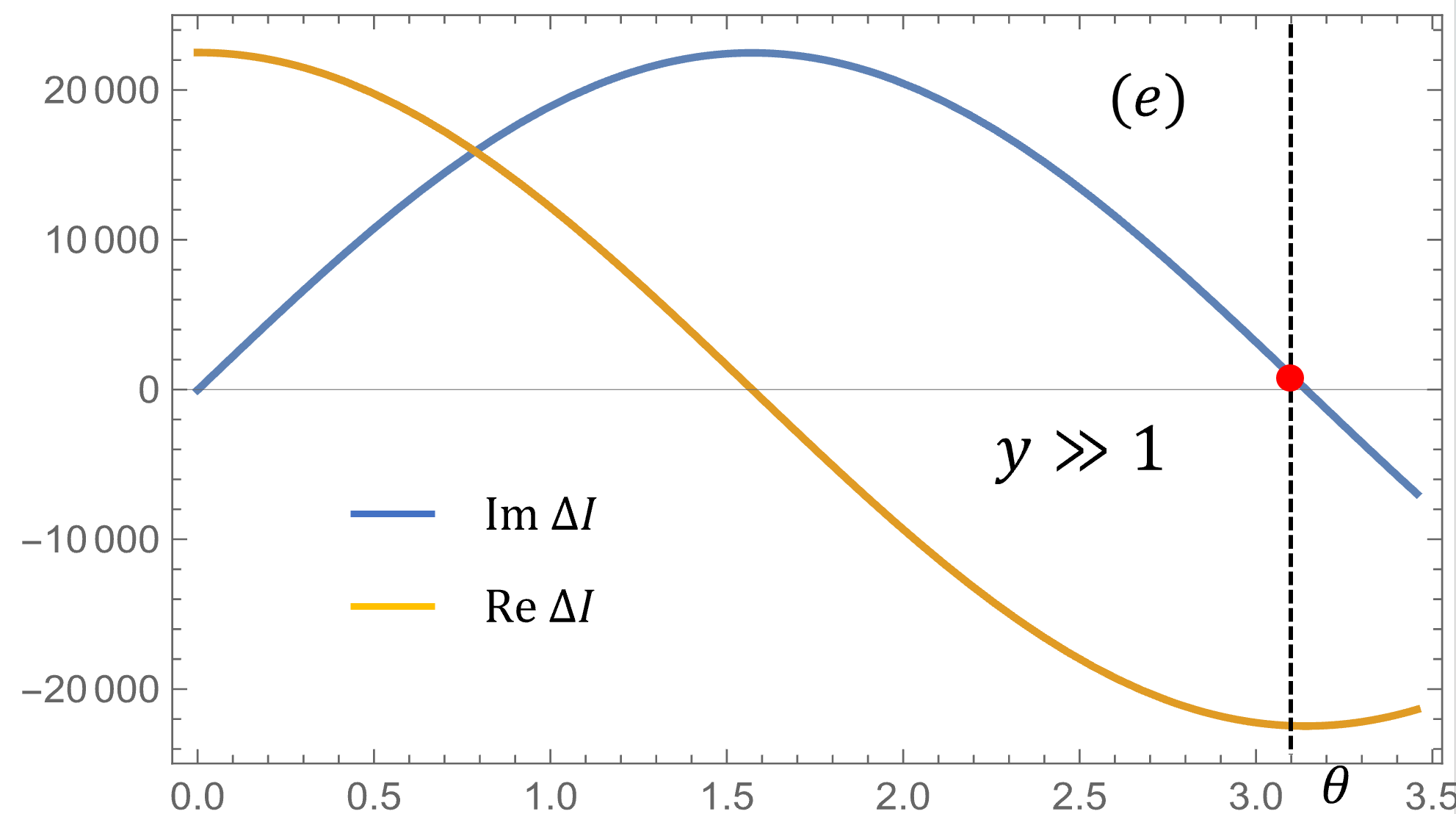}}
    \caption{Plots of $\text{Im} \Delta I$ (blue) and $\text{Re} \Delta I$ (yellow) featuring a Stokes phenomenon at $\theta=\pi$ (red dot) for the saddle-points $\zeta^*_{s,s'}$. The re-scaled temperature is of order 1 for $y=1.8$ and $|\lambda| =10^{-2}$ for (a)-(d); at low temperatures $y\gg 1$, the plots for distinct $s,s'$ become almost identical, shown in (e) for $y=150$ and $\lambda = 10^{-2}$. }
    \label{fig:Stokes_pheno}
\end{figure}

We now apply the analysis to the integral representation (\ref{eq:integral_rep}), with $Z_{\text{CFT}}$ approximated by the toy model (\ref{eq:toy_model}). Along the positive real axis $\lambda = |\lambda|>0$ and for temperatures satisfying $\tau_2 > \tau^c_2 = \sqrt{\pi c\lambda/3}$, it can be checked that the integration contour $\mathcal{H}_+$ decomposes into only the Lefschetz thimble $\mathcal{J}_p$ through the physical saddle-point  $\zeta^*_p = (\zeta^*_1,\zeta^*_2)$ in the class (\ref{eq:saddle_1}): 
\be 
\zeta^*_1 = 0,\;\;\zeta^*_2 = \sqrt{\frac{3\tau^2_2-\pi c \lambda}{3-\pi c \lambda}} \approx \tau_2+...
\ee
In fact for $\lambda>0$, the integration contour $\mathcal{H}_+$ actually coincides with $\mathcal{J}_p$, giving: 
\be\label{eq:PF_positive}
\mathcal{H}_+ = \mathcal{J}_p\;\;\rightarrow \;\;Z(\tau,\bar{\tau},\lambda) = Z_p(\tau,\bar{\tau},\lambda) \sim e^{-I_{TM}(\zeta^*_p)},\;\;I_{TM}(\zeta^*_p)= \frac{2}{3\lambda}\sqrt{\left(3\tau_2^2-\pi c \lambda\right)\left(3-\pi c \lambda\right)}-\frac{2\tau_2}{\lambda}
\ee
In particular, the contributions from all the other saddle-points are absent for positive $\lambda$. Notice that had they been present, they will amount to exponentially larger contributions of the order $e^{2\tau_2/|\lambda|}$ or $e^{4\tau_2/|\lambda|}$. As $\lambda = e^{i\theta}|\lambda|$ rotates in the complex $\lambda$-plane while kept at small modulus $|\lambda|<1$, the Lefschetz thimbles deform accordingly. Our focus is the Stokes' phenomenon associated with the saddle-points from (\ref{eq:saddle_2}). Based on the general discussion, this simply requires us to identify the occurrences of the condition (\ref{eq:Stokes_pheno}) on the complex $\lambda$-plane. Let us label by $\zeta^*_{s,s'}$ on of the saddle-points from (\ref{eq:saddle_2}): 
\be 
\zeta^*_{s,s',s''} = \left(\zeta^*_1,\zeta^*_2\right),\;\;\zeta^*_1 =  s''\sqrt{\frac{6\tau_2^2+ 2\pi s c\tau_2 \lambda}{2\pi c\lambda}},\;\;\zeta^*_2 = \ri\tau_2 s'\sqrt{\frac{3}{\pi c\lambda}},\;\;s,s',s''= \pm 1
\ee
The saddle-contribution from $\zeta^*_{s,s'}$ is given by: 
\be
Z_{s,s',s''}(\tau,\bar{\tau},\lambda)\sim e^{-I_{TM}\left(\zeta^*_{s,s',s''}\right)},\;\;\; I_{TM} \left(\zeta^*_{s,s',s''}\right) = -\ri\sqrt{\frac{4\pi c}{3\lambda}}B - \frac{2\tau_2}{\lambda},\;\;B = s'\tau_2+ss'
\ee
Since it does not depend on $s''$, we omit the index $s''$ from now on. The condition (\ref{eq:Stokes_pheno}) for the Stokes' phenomenon associated with the saddle-point $\zeta^*_{s,s'}$ amounts to that: 
\be
\text{Im}\;\Delta I = 0,\;\;\;\text{Re}\;\Delta I<0
\ee
The action difference $\Delta I$ at the leading orders in small $\lambda$ can be written as: 
\bea
\Delta I &=& I_{TM}\left(\zeta^*_p\right) - I_{TM}\left(\zeta^*_{s,s'}\right) = \lambda^{-1}\sqrt{\tau_2^2-\frac{\pi c\lambda}{3}}+i \sqrt{\frac{4\pi c }{3\lambda}} B +...\nonumber\\
&=& \frac{\tau^c_2}{|\lambda|} \left(e^{-i\theta}\sqrt{y^2-e^{i\theta}}+2i B e^{-i\theta/2}\right)+...
\eea
for the complex $\lambda = e^{i\theta}|\lambda|$ and the re-scaled temperature $y$ defined by $\tau_2 =  y\;\tau^c_2$. The condition (\ref{eq:Stokes_pheno}) for the Stokes' phenomenon involving $\zeta^*_{s,s'}$ can therefore be written explicitly in terms of phase $\theta$: 
\bea\label{eq:Stokes_pheno_2} 
&&A \sin{\left(\Theta+\theta\right)}-2B \cos{\left(\frac{\theta}{2}\right)}=0,\;\; A\cos{\left(\Theta+\theta\right)}+2B \sin{\left(\frac{\theta}{2}\right)}<0\nonumber\\
&&A=\sqrt{y^4-2y^2 \cos{\theta}+1},\;\;\tan{\Theta} = \frac{\sin{\theta}}{y^2-\cos{\theta}}
\eea
It can be checked that for $y>1$ and starting from $\theta =0$, the Stokes' phenomena occurs at $\theta=\pi$ for any choice of $s,s'$. For illustration in Figure (\ref{fig:Stokes_pheno}) We plot $\text{Im}\; \Delta I$ and $\text{Re}\; \Delta I$ as functions of $\theta$, both for low temperatures $y\gg 1$ or near the critical temperature $y\sim \mathcal{O}(1)$.

We conclude as $\lambda$ analytically continue from $\lambda>0$, it receives non-perturbative contributions from the saddle-points in (\ref{eq:saddle_2}) when $\lambda\to -|\lambda|<0$. We can therefore write the continued partition function in terms of the positive parameter $|\lambda|$ as: 
\bea\label{eq:PF_negative}
Z(\tau,\bar{\tau},-|\lambda|) &=& Z_p(\tau,\bar{\tau},-|\lambda|) + \sum_{s,s'} Z_{s,s'}(\tau,\bar{\tau},-|\lambda|) \nonumber\\
&\sim & e^{-\frac{2\tau_2}{|\lambda|}+\frac{2}{3|\lambda|}\sqrt{\left(3\tau_2^2+\pi c |\lambda|\right)\left(3+\pi c |\lambda|\right)}} + e^{-\frac{2\tau_2}{|\lambda|}+\sqrt{\frac{4\pi c}{3|\lambda|}}+...} +...
\eea
To clarify, in writing the non-perturbative corrections $Z_{s,s'}$ in the second line we have only kept the contributions from the saddle-points with $ss'=1$ because they dominate over the others at the order of $\mathcal{O}(|\lambda|^{-1/2})$ in the exponent. In addition, we also ignored the sub-leading corrections of the same order to the coefficient of $\tau_2$ in the exponent. Notice that while (\ref{eq:PF_positive}) as a function of the inverse temperature $\tau_2$ has a physical singularity at $\tau^c_2>0$, the expression (\ref{eq:PF_negative}) is now singularity-free for real and positive $\tau_2$. So we can use this expression to obtain an analytic continuation of $Z(\tau,\bar{\tau},-|\lambda|)$ above the critical temperature $T=1/\tau^c_2$, making it well-defined for all physical $\tau_2>0$. 

We make a few comments before concluding the section. What we have performed is one particular procedure for analytic continuing past $\tau^c_2$ at negative $\lambda$. An alternative approach could proceed by first finding the correct partition function for $\tau_2 < \tau^c_2$ at positive $\lambda$, and then rotate $\lambda$ to the negative real-axis. This could in principle produce different (and possibly more physical) results, but the first step involves resolving the Hagedorn-like singularity at $\tau^c_2$ \cite{PhysRevLett.25.895,Fubini:1969qb,Hagedorn:1965st}. This is an extremely interestintg endeavor that is beyond the current scope of the work. Motivated by the understanding in string theory \cite{Sathiapalan:1986db,Kogan:1987jd,OBrien:1987kzw,ATICK1988291}, it possibly involves studying the winding mode contributions, or even contributions from the target space of non-torus topology, in the flat JT-gravity or non-critical string theory formulation of the $T\bar{T}$-deformation \cite{Dubovsky:2018bmo,Callebaut:2019omt,Hashimoto:2019wct}, from which the integral representation (\ref{eq:integral_rep}) is derived but did not incorporate. We leave this for future studies. We also remark that in principle there is also the non-perturbative contribution from the remaining saddle of (\ref{eq:saddle_1}). We did not consider it because its contribution $\sim e^{-4\tau_2/|\lambda|}$ is more suppressed at $\lambda<0$ than those of (\ref{eq:saddle_2}); and related to this its resurgence effect is too weak to be observable in the perturbative expansion. It is also interesting to revisit this in future works. 

\section{Generalised Richardson transform}
\label{sc:Rich}

Given the asymptotic behavior of a function $f(n)$ at large $n$
\begin{equation}
    f(n) \sim f_0 + \frac{f_k}{n^k} +\mc{O}(n^{-k-1}),\quad n\rightarrow \infty
\end{equation}
with $k\in\IN$,
the standard method to remove the $1/n^k$ term, even if $f_k$ is not known, so that we approach $f_0$ faster as $n\rightarrow \infty$ is the Richardson transform of the $k$-th order
\begin{equation}
\label{eq:Rkf}
    R_k[f](n) = \frac{n^kf(n)-(n-s)^kf(n-s)}{n^k - (n-s)^k},
\end{equation}
as 
\begin{equation}
    R_k[f](n) \sim f_0 + \mc{O}(n^{-k-1}),\quad n\rightarrow \infty.
\end{equation}
See for instance \cite{Bender78}.

One can actually show that this works for any $k\in \IQ$.
For instance, if 
\begin{equation}
    f(n) \sim f_0 + \frac{f_{1/2}}{n^{1/2}} + \mc{O}(1/n),\quad n\rightarrow \infty,
\end{equation}
then
\begin{equation}
    R_{1/2}[f](n)\sim f_0 +\mc{O}(1/n),\quad n\rightarrow\infty,
\end{equation}
where $R_k[f](n)$ is \eqref{eq:Rkf} with $k=1/2$.
We will call this the generalised Richardson transform.

\end{appendix}

\end{document}